\makeatletter
\def\input@path{{sections/}{algos/}{figs/}{figs/code-samples/}}
\makeatother

\documentclass[sigplan,screen,balance,nonacm]{acmart}

\usepackage{microtype}
\usepackage{bm}
\usepackage{listings}
\usepackage[boxed]{algorithm2e}
\usepackage{tikz}
\usepackage{subcaption}
\usepackage{adjustbox}
\usepackage{siunitx}
\usepackage{enumitem}

\usetikzlibrary{quantikz2}
\setlist[itemize]{noitemsep, topsep=0pt}

\graphicspath{ {./figs/} }

\lstdefinestyle{ir}{language=LLVM,morestring=[b]',morekeywords={rev,irrev,qbundle,qubit,qbundle,bitbundle,i1,f64,qbprep,qbdiscard,qbdiscardz,qbpack,qbunpack,bitpack,bitunpack,qbtrans,qbmeas,qbphase,by,in,phases,func,adj,pred,lambda,as,func_const,func_adj,func_pred,call_indirect,if,arith.const,calc,array,callable,qalloc,qfree,qfreez,measure,gate,callable_metadata,callable_create,callable_adj,callable_ctrl,callable_invoke,arrpack,arrunpack,scf.if,else,scf.yield,qwerty.func_const,qwerty.call_indirect},basicstyle=\ttfamily,emphstyle=\bfseries\color{orange!70!black},keepspaces=true,stringstyle=\color{blue},showstringspaces=false,aboveskip=0pt,belowskip=0pt}
\lstdefinestyle{irs}{style=ir,basicstyle=\ttfamily\small}
\lstdefinestyle{irxs}{style=ir,basicstyle=\ttfamily\footnotesize}
\lstdefinestyle{base}{language=Python,emph={@qpu,@classical},morekeywords={kern,cmap,qmap,qubit,bit,basis,std,pm,ij,id,fourier,exp,discard,xor_reduce,and_reduce,sign,inplace,xor,measure,flip,cfunc,func,rev_func},basicstyle=\ttfamily,emphstyle=\bfseries\color{orange!70!black},keepspaces=true,stringstyle=\color{blue},showstringspaces=false,aboveskip=0pt,belowskip=0pt}
\lstdefinestyle{s}{style=base,basicstyle=\ttfamily\small}
\lstdefinestyle{xs}{style=base,basicstyle=\ttfamily\footnotesize}
\lstdefinestyle{sqs}{style=base,basicstyle=\ttfamily\small,language=QSharp}
\lstdefinestyle{num}{style=base,frame=single,numbers=left,numberstyle=\sffamily\footnotesize\color{gray},numbersep=4.5pt}
\lstdefinestyle{nums}{style=num,basicstyle=\ttfamily\small}
\lstdefinestyle{numxs}{style=num,basicstyle=\ttfamily\footnotesize}
\lstdefinestyle{numxxs}{style=num,basicstyle=\ttfamily\scriptsize}
\lstdefinestyle{numxxxs}{style=num,basicstyle=\ttfamily\tiny}
\DeclareMathOperator{\vspan}{span}
\DeclareMathOperator{\vdim}{dim}
\DeclareMathOperator{\vprim}{prim}

\newcommand{\toktensor}{\mathbin{\texttt{+}}}

\newcommand{\varbe}{\mathrm{be}}
\newcommand{\varbein}{\varbe_{\mathrm{in}}}
\newcommand{\varbeout}{\varbe_{\mathrm{out}}}
\newcommand{\varbin}{b_{\mathrm{in}}}
\newcommand{\varbout}{b_{\mathrm{out}}}
\newcommand{\varbbig}{b_{\mathrm{big}}}
\newcommand{\varbsmall}{b_{\mathrm{small}}}
\newcommand{\varprim}{\mathrm{prim}}
\newcommand{\hilbertspace}{\mathcal{H}_2}

\newcommand{\algocomment}[1]{\tcp*[f]{#1}}
\newenvironment{algo}{
    \IncMargin{5pt}
    \begin{algorithm}
    \DontPrintSemicolon
    \SetAlgoNoEnd
    \LinesNumbered
    \SetKwInput{Input}{Input}\SetKwInput{Output}{Output}
    \SetKw{Continue}{continue}
}{
    \end{algorithm}
}

\newcommand{\compilername}{\textsc{Asdf}}
\newcommand{\ofsup}{}
\newcommand{\ofmain}{}

\providecommand{\lnbvsimpleclassical}{5}
\providecommand{\lnbvsimpleclassicalret}{7}

\providecommand{\lnbvsimplesecretbegin}{5}
\providecommand{\lnbvsimplesecretend}{6}

\providecommand{\lnteleportevilbegin}{10}
\providecommand{\lnteleportevilend}{11}

\begin{document}

\title{\compilername{}: A Compiler for Qwerty, a Basis-Oriented Quantum Programming Language}

\author{Austin J. Adams}
\orcid{0000-0002-3179-8735}
\affiliation{%
  \institution{Georgia Institute of Technology}
  \city{Atlanta, GA}
  \country{USA}
}
\email{aja@gatech.edu}
\authornote{Corresponding author}

\author{Sharjeel Khan}
\orcid{0000-0002-4563-4619}
\affiliation{%
  \institution{Georgia Institute of Technology}
  \city{Atlanta, GA}
  \country{USA}
}
\email{smkhan@gatech.edu}

\author{Arjun S. Bhamra}
\orcid{0000-0002-1557-5988}
\affiliation{%
  \institution{Georgia Institute of Technology}
  \city{Atlanta, GA}
  \country{USA}
}
\email{abhamra@gatech.edu}

\author{Ryan R. Abusaada}
\orcid{0009-0008-5069-0296}
\affiliation{%
  \institution{Georgia Institute of Technology}
  \city{Atlanta, GA}
  \country{USA}
}
\email{rabusaada6@gatech.edu}

\author{Anthony M. Cabrera}
\orcid{0000-0002-6561-0382}
\affiliation{%
  \institution{Oak Ridge National Laboratory}
  \city{Oak Ridge, TN}
  \country{USA}
}
\email{cabreraam@ornl.gov}

\author{Cameron C. Hoechst}
\orcid{0009-0008-1574-3700}
\affiliation{%
  \institution{Georgia Institute of Technology}
  \city{Atlanta, GA}
  \country{USA}
}
\email{choechst3@gatech.edu}

\author{Travis S. Humble}
\orcid{0000-0002-9449-0498}
\affiliation{%
  \institution{Oak Ridge National Laboratory}
  \city{Oak Ridge, TN}
  \country{USA}
}
\email{humblets@ornl.gov}

\author{Jeffrey S. Young}
\orcid{0000-0001-9841-4057}
\affiliation{%
  \institution{Georgia Institute of Technology}
  \city{Atlanta, GA}
  \country{USA}
}
\email{jyoung9@gatech.edu}

\author{Thomas M. Conte}
\orcid{0000-0001-7037-2377}
\affiliation{%
  \institution{Georgia Institute of Technology}
  \city{Atlanta, GA}
  \country{USA}
}
\email{conte@gatech.edu}

\renewcommand{\shortauthors}{Adams, Khan, Bhamra, Abusaada, Cabrera, Hoechst, Humble, Young, and Conte}

\begin{abstract}
Qwerty is a high-level quantum programming language built on bases and functions rather than circuits. This new paradigm introduces new challenges in compilation, namely synthesizing circuits from basis translations and automatically specializing adjoint or predicated forms of functions. This paper presents \compilername{}, an open-source compiler for Qwerty that answers these challenges in compiling basis-oriented languages. Enabled with a novel high-level quantum IR implemented in the MLIR framework, our compiler produces OpenQASM~3 or QIR for either simulation or execution on hardware. Our compiler is evaluated by comparing the fault-tolerant resource requirements of generated circuits with other compilers, finding that \compilername{} produces circuits with comparable cost to prior circuit-oriented compilers.
\end{abstract}

\maketitle

\section{Introduction}
Qwerty is a recently proposed, Python-embedded, high-level quantum programming language built on quantum basis translations rather than quantum gate circuitry~\cite{adams_qwerty_2024}.  Although the language provides a valuable scaffold for reasoning about quantum algorithms, its compilation is challenging due to several fundamental constructs. Chief among these constructs is its \textit{basis translation}, a novel transformation that shifts qubits from one basis to another, which the compiler must synthesize as a quantum circuit. In addition, functions in Qwerty may be invoked backward or in a predicated form, requiring the compiler to generate \textit{function specializations}. However, modern quantum compilers cannot compile basis translations and their synthesis of function specializations is more constrained.
Moreover, current intermediate representations (IRs) for quantum programs are ill-suited to represent Qwerty programs due to the semantic gap between Qwerty and quantum circuits. 

This paper presents \textit{\textbf{\compilername{}}}\footnote{
\,``\compilername{}'' is a keyboard pun on ``Qwerty.''}, the first compiler for
Qwerty's new basis-oriented quantum programming paradigm. \compilername{} spans
the chasm between high-level code expressed with functions and bases and
efficient low-level circuits by leveraging a novel quantum IR design implemented
in the MLIR framework. We evaluate the efficiency of circuitry produced by
\compilername{} in industry-standard formats such as OpenQASM 3 and QIR across a
suite of well-known quantum algorithms written in Qwerty, finding
\compilername{}'s output competitive with handwritten assembly.

The backbone of \compilername{} is a runtime that retrieves the Python AST
(abstract syntax tree)~\cite{ast} and type checks it to enforce quantum
semantics and prevent violation of the postulates of quantum computation.
Type checking Qwerty's quantum constructs could result in exponential runtime,
but we present an efficient type checker that uses quantum basis factorization
to avoid this pitfall.
Because of the nature of quantum algorithm expression in Qwerty, the Qwerty AST
consists of a pipeline of function calls, which \compilername{} must rebuild
into a more imperative structure for downstream compatibility. We tackle this
challenge by integrating a custom IR dialect into MLIR such that function
reversal and predication are
computed in full generality.
Aggressive inlining aiming
to linearize the computation lowers this IR into a quantum dataflow description
of the quantum circuit, where peephole
optimizations are applied to gate subgraphs to reduce the complexity and
improve the quality of the output. An imperative circuit description can be
easily generated from this dataflow
graph; two formats are implemented.
One is standard OpenQASM3 that can be sent directly to IBM for
execution~\cite{cross_openqasm_2022,healy_design_2024}. The other is QIR, which
can be input to QIR-EE to target several different hardware and simulation
platforms~\cite{qir,wong_cross-platform_2024}.
The source code for \compilername{} is available at \url{https://github.com/gt-tinker/qwerty}.

This paper makes the following contributions:
\begin{itemize}[topsep=0pt]
    \item Type checking basis translations efficiently to avoid exponential runtime;
    \item \textit{Qwerty IR}, an IR designed both to represent the unique structure of Qwerty programs and to flatten easily into a quantum circuit;
    \item Reversing or predicating quantum basic blocks effectively;
    \item Novel techniques for synthesizing circuits for basis translations with minimal overhead; and
    \item A demonstration that the compiled code is of equal quality to handwritten quantum circuits in Qiskit, Quipper, and Q\# for several well-known quantum algorithms~\cite{Qiskit,green_quipper_2013,svore_q_2018}.
\end{itemize}

\noindent The following section presents a brief introduction to quantum notation followed by an overview of Qwerty. Afterward, an overview of the \compilername{} compiler is provided and each compilation phase is discussed. Empirical comparisons of the compiler's output to hand-built circuits are then presented, followed by a discussion of related work and overall conclusions.

\section{Background and Motivation}
\subsection{Quantum Computing Notation}\label{sec:quantum-notation}
For a thorough introduction to quantum computing and its notation, we direct
readers to Chapters 1 and 2, respectively, of Nielsen and
Chuang~\cite{nielsen_quantum_2010}. However, some definitions are included here for clarity. We define the \textit{Z eigenbasis} as $\ket{0}\!/\ket{1}$; the
\textit{X eigenbasis} as $\ket{+}\!/\ket{-}$ (or equivalently $\ket{p}\!/\ket{m}$);
and the \textit{Y eigenbasis} as
$\ket{i}\!/\ket{j}$, where $\ket{i} = \frac{1}{\sqrt{2}}(\ket{0} + i\ket{1})$ and
$\ket{j} = \frac{1}{\sqrt{2}}(\ket{0} - i\ket{1})$~\cite{nielsen_quantum_2010}.
We call the first vector in each of these pairs the \textit{plus eigenstate}
(since it has eigenvalue $+1$ for the respective Pauli) and the second vector
the \textit{minus eigenstate} (since it has eigenvalue $-1$ for the respective
Pauli).
The relative phase shift gate $P(\theta)$ is defined as
$\proj{0} + e^{i\theta}\proj{1}$~\cite{smith_practical_2017}.

\begin{figure}
\centering
\begin{lstlisting}[style=numxs,framexleftmargin=4mm,xleftmargin=6mm,linewidth=0.98\linewidth]
import sys
from qwerty import *

def bv(secret_str):
  @classical[N](secret_str)
  def f(secret_str: bit[N], x: bit[N]) -> bit:
    return (secret_str & x).xor_reduce()

  @qpu[N](f)
  def kernel(f: cfunc[N,1]) -> bit[N]:
    return 'p'[N] | f.sign \
                  | pm[N] >> std[N] \
                  | std[N].measure

  return kernel()

secret_str = bit.from_str(sys.argv[1])
print(bv(secret_str))
\end{lstlisting}
\caption{Bernstein--Vazirani expressed in Qwerty~\cite{adams_qwerty_2024}\label{fig:bv_simple}}
\end{figure}

\subsection{An Introduction to Qwerty\label{sec:qwerty-intro}}
Although Qwerty is defined in prior work~\cite{adams_qwerty_2024}, a reference
for relevant Qwerty features is included here for convenience, since it affects
type checking (e.g., Section~\ref{sec:span-equiv-checking}) and influences the design of
our basis-oriented IR (Section~\ref{sec:qwerty-dialect}).

As discussed above, Qwerty is a \textit{basis-oriented} quantum programming
language~\cite{adams_qwerty_2024}. In contrast with prior quantum programming
languages, Qwerty programmers write quantum programs using 
primitives defined in terms of orthonormal bases or classical logic, not
by choosing from a library of quantum gates. (Fig.~\ref{fig:bv_simple} is an
example of a Qwerty program~\cite{adams_qwerty_2024}.)

Every basis in Qwerty is grounded in four \textit{primitive bases}: \lstinline!std! (the $Z$
eigenbasis), \lstinline!pm! (the $X$ eigenbasis), \lstinline!ij! (the $Y$
eigenbasis), and \lstinline!fourier[N]! (the \lstinline!N!-qubit Fourier
basis~\cite[\S5.1]{nielsen_quantum_2010}). Each primitive basis can be used in
a Qwerty \textit{basis expression}: the tensor product of two bases is written
$b_1\texttt{ + }b_2$, and the $N$-fold tensor product is written
$b\texttt{[}N\texttt{]}$.
Basis expressions may also contain
\textit{basis literals}, written
$\texttt{\{}\mathrm{bv}_1\!\texttt{,}\mathrm{bv}_2\!\texttt{,}\ldots\!\texttt{,}\mathrm{bv}_m\texttt{\}}$
where $m > 0$. Each \textit{basis vector} $\mathrm{bv}$ is a \textit{qubit
literal} with an optional complex unit scalar phase factor (written
\lstinline!bv@theta! to evoke \lstinline!bv!$\bm{\circlearrowleft}$\lstinline!theta!). Qubit literals resemble string literals
(e.g., \lstinline!'1010'!) and are sequences of {\color{blue}\texttt{p}},
{\color{blue}\texttt{m}}, {\color{blue}\texttt{i}}, {\color{blue}\texttt{j}},
{\color{blue}\texttt{0}}, and {\color{blue}\texttt{1}}, each \textit{position} of this sequence
corresponding to the vectors in Section~\ref{sec:quantum-notation}. The
\textit{dimension} of a qubit literal, basis vector, or basis is its number of
qubits. The \textit{eigenbits} of a basis vector is a bitstring whose length is
the dimension of the vector; each bit is set if and only if the corresponding
position is a minus eigenstate. In a well-typed basis literal, all eigenbits must be
distinct, all vector dimensions must be equal, and all positions of all vectors
must have the same primitive basis (defined as the non-\lstinline!fourier! primitive basis
the respective position belongs to). A \textit{canon form} of a basis is a
sequence (tensor product) of \textit{basis elements}, each of which is either a
basis literal or a \textit{built-in basis}, an $N$-qubit primitive basis (e.g.,
\lstinline!pm[4]!). Any Qwerty basis can be written in canon form.

The core computational primitive of Qwerty is the \textit{basis
translation}~\cite{adams_qwerty_2024}. Given an input basis $\varbin$
and an output basis $\varbout$, a basis translation $\varbin \:\texttt{>{}>}\:
\varbout$ performs the following operation (where the vectors $\vert b^{(i)} \rangle$ are the basis vectors of a basis $b$):
\begin{gather*}
    \alpha_1\ket{\varbin^{(1)}}
    + \alpha_2\ket{\varbin^{(2)}}
    + \cdots
    + \alpha_n\ket{\varbin^{(n)}}
    {\color{darkgray} {}+ \ket{\varbin^\perp}} \\
    \bm{\Downarrow} \\
    \alpha_1\ket{\varbout^{(1)}}
    + \alpha_2\ket{\varbout^{(2)}}
    + \cdots
    + \alpha_n\ket{\varbout^{(n)}}
    {\color{darkgray} {}+ \ket{\varbin^\perp}}
\end{gather*}
That is, a basis translation preserves state amplitudes but translates the
basis vectors from $\varbin$ to $\varbout$; amplitudes are preserved because
all possible translations are unitary. This unitarity promise is accomplished
by all Qwerty bases being orthonormal and Qwerty requiring that both
bases in a basis translation span the same space: $\vspan(\varbin) = \vspan(\varbout)$.

For example, the basis translation \lstinline!{'01','10'}! \lstinline!>>! \lstinline!{'10',! \lstinline!'01'}!
swaps two qubits, since it performs the following operation:
\begin{gather*}
    {\color{darkgray} \beta_1\ket{00} +{}}
      \alpha_1\bm{\ket{01}}
    + \alpha_2\bm{\ket{10}}
    {\color{darkgray} {}+ \beta_2\ket{11}} \\
    \bm{\Downarrow} \\
    {\color{darkgray} \beta_1\ket{00} +{}}
      \alpha_1\bm{\ket{10}}
    + \alpha_2\bm{\ket{01}}
    {\color{darkgray} {}+ \beta_2\ket{11}}
\end{gather*}
This example is identical to a SWAP gate in circuit-level quantum programming.

\begin{figure*}
    \centering
    \includegraphics[width=0.8\linewidth]{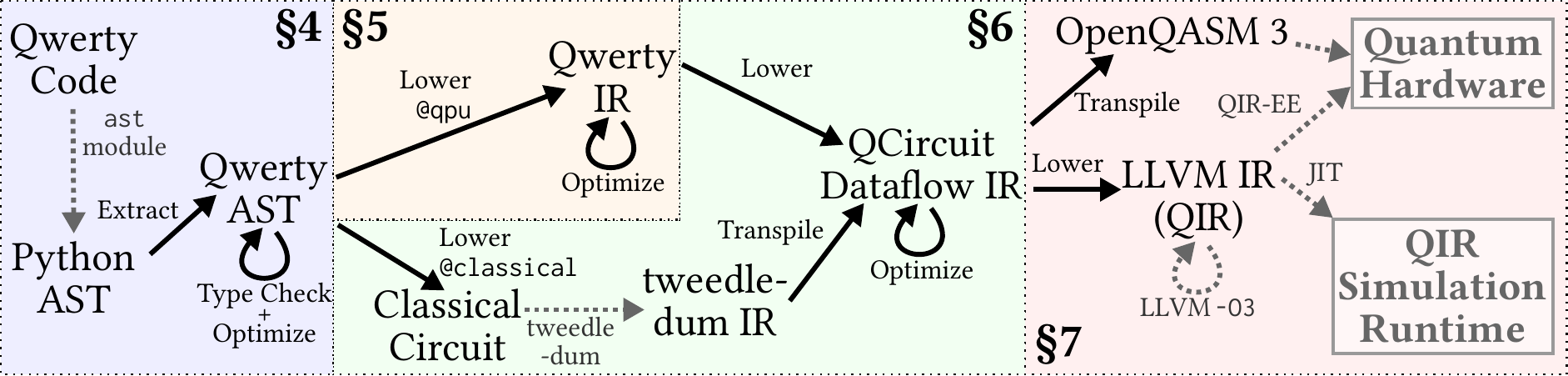}
    \Description{A cartoon showing our compilation flow.}
    \caption{Compilation flow for our Qwerty compiler. Solid arrows are our work.}
    \label{fig:compilation-flow}
\end{figure*}

In quantum algorithms, it is common to run a sequence of quantum operations
backwards or in a proper subspace.
Qwerty makes the intuitive choice to let these reversible (or predicatable) sequences of quantum operations be
functions that take \lstinline!qubit!s and return \lstinline!qubit!s. All quantum
operations in these reversible functions must themselves be reversible; such
functions are designated by having the type $T_1 \xrightarrow{\text{rev}} T_2$
\cite[\S A]{adams_qwerty_2024}. A version of a reversible function
\lstinline!f! than runs backwards (i.e., an adjointed form) is written in
Qwerty as \lstinline!~f!. A version which runs only when the span of a basis
\lstinline!b! contains the state of some additional $(\vdim{}$\lstinline!b!$)$
input qubits (i.e., a predicated form) is written as \lstinline!b & f!. Both
take an operand of type
\lstinline!qubit[N]!${}\xrightarrow{\text{rev}}{}$\lstinline!qubit[N]!.

\subsection{Motivation: Challenges in Compiling Qwerty}
Programming in Qwerty requires a compiler that meets the following challenges:
\begin{enumerate}
    \item \textbf{Handling bases:}
          The basis-oriented primitives in Qwerty must be \textit{efficiently}
          type checked and lowered to gates, the standard quantum program
          representation expected by quantum compiler backends.
    \item \textbf{Linearizing code:}
          The functional constructs in Qwerty code such as indirect calls,
          adjoint calls, and predicated calls must be converted into
          straight-line code as much as possible because functional features may not
          be well-supported in quantum hardware.
    \item \textbf{Integration with standard tools:}
          For Qwerty to be practical, a compiler should produce quantum circuits in a
          format recognized by industry-standard tools such as QIR or OpenQASM.
\end{enumerate}
The remainder of this paper is dedicated to how the design of \compilername{}
addresses these challenges.

\section{Overview of the \compilername{} Compiler}
\label{ref:overview_comp}

Fig.~\ref{fig:compilation-flow} illustrates the compilation process for a
Qwerty program in the \compilername{} compiler. The figure is divided into regions of steps
that correspond to the following sections of this paper:
 \textbf{Section~\ref{sec:ast}} covers the generation and transformations done on
      the abstract syntax tree (AST), notably including type checking.
\textbf{Section~\ref{sec:qwerty-dialect}} defines the high-level IR used to
      represent and optimize Qwerty programs (\textit{Qwerty IR}), implemented as a dialect in MLIR.
\textbf{Section~\ref{sec:quantum-dialect}} describes how Qwerty IR is lowered to a
      lower-level gate-based IR. This includes synthesizing circuits from basis
      translations and generating specializations (reversed and predicated
      forms) of functions.  Finally, 
\textbf{Section~\ref{sec:targeting}} discusses how \compilername{} generates
      QIR and OpenQASM~3.
\\

\noindent\textbf{Frameworks used in implementation:}
The implementation of \compilername{} involves two frameworks commonly used for quantum compiler implementation, MLIR and QIR.

\noindent\textbf{\textit{MLIR:}}
The Multi-Level IR compiler framework, commonly called MLIR, is a popular and
extensible compiler infrastructure within the LLVM project, and its design
lends itself to supporting domain specific languages and IRs~\cite{lattner_mlir_2021}.
MLIR supports user-defined \textit{dialects} containing reusable types,
\textit{attributes} (metadata known at compile time), and \textit{ops}
(instructions).
Common optimizations ship with MLIR, including function inlining and
\textit{canonicalization}; the latter simplifies IR to better enable optimizations (e.g., through constant folding and dead code elimination).
MLIR dialects exist for representing quantum programs using gates or
domain-specific ops~\cite{mccaskey_mlir_2021,peduri_qssa_2022,ittah_qiro_2022,The_CUDA-Q_development_team_CUDA-Q, Ittah_Catalyst_a_Python_2024,healy_design_2024}.

\noindent\textit{\textbf{QIR:}}
\compilername{} is designed to produce Quantum Intermediate
Representation (QIR), an LLVM IR--based quantum IR designed to be a target for
high-level quantum programming languages such as Microsoft's
Q\#~\cite{lubinski_advancing_2022,QIRAlliance,svore_q_2018}. To respect the
capabilities of different hardware, QIR has different \textit{profiles}, including
the \textit{Base Profile}~\cite{Wernli_Heim_2022}, which allows only forward unconditional
branching and bans dynamic qubit allocation, and the \textit{Unrestricted
Profile}~\cite{qsharp_unrestricted}, which permits the complete library of QIR intrinsics and full
generality of LLVM IR.

\section{Abstract Syntax Tree}\label{sec:ast}
\noindent\textbf{AST generation:}
Compilation in \compilername{} begins similarly to
to Python DSLs for classical accelerators such as
Intrepyyd~\cite{zhou_intrepydd_2020} or Triton~\cite{tillet_triton_2019}: \compilername{} retrieves the Python AST (itself a Python data structure) for
\lstinline!@qpu! kernels and \lstinline!@classical! functions using the Python
standard library \cite{ast}.
(Thus, the \compilername{} compiler and runtime are distributed as a Python package, so no changes to the Python interpreter itself are needed.)
While easy to manipulate, the Python AST is untyped and
intended for Python programs, not Qwerty programs, so \compilername{}
converts it to a typed Qwerty AST by recognizing the patterns in the AST
formed by Qwerty syntax.

\noindent\textbf{AST expansion:}
\compilername{} then preprocesses the AST in advance of type checking.
Qwerty supports \textit{dimension variables}, which allow defining a function
polymorphically with respect to some integer dimension (e.g., input problem
size).
A Qwerty compiler should infer dimension
variables based on the types of captures when possible --- for example, \compilername{} infers \lstinline!N! from the length of the captured secret bitstring
on lines~\lnbvsimplesecretbegin{}-\lnbvsimplesecretend{} in
Fig.~\ref{fig:bv_simple}. Once dimension variables are determined, whether by
inference or by the programmer explicitly providing them, \compilername{}
performs \textit{expansion} on the AST, replacing all expressions involving
dimension variables with constants. This includes unrolling loops and expanding
\textit{expr}\lstinline![N]! to
\textit{expr} \lstinline!+! \textit{expr} \lstinline!+! $\cdots$ \lstinline!+!
\textit{expr}.

\noindent\textbf{AST type checking:}
After expansion, \compilername{} type checks the AST. The \compilername{} type checker enforces linear types for qubits: any quantum
value must be used exactly once and cannot be discarded~\cite{adams_qwerty_2024,selinger_lambda_2006,yuan_twist_2022,paykin_qwire_2017}. Type checking also verifies that functions marked as reversible call
only other reversible functions and have no classical conditional expressions~\cite{adams_qwerty_2024,svore_q_2018,singhal_q_2022}.
The most novel parts of the \compilername{} type checker involve validation of bases and
basis translations. For example, the compiler must verify that each basis
literal truly represents an orthonormal basis by checking the conditions on
basis literals enumerated in Section~\ref{sec:qwerty-intro}.
Type checking for a basis translation is more involved, as described below.

\subsection{Span Equivalence Checking}\label{sec:span-equiv-checking}

A basis translation $\varbin$\lstinline! >> !$\varbout$ passes type checking if
$\varbin$ and $\varbout$ pass type checking and $\vspan(\varbin) =
\vspan(\varbout)$, that is, if both (valid) bases span exactly the same
subspace of qubit states \cite[\S A.2]{adams_qwerty_2024}. Henceforth, we call
the problem of determining that $\vspan(\varbin) = \vspan(\varbout)$
\textit{span equivalence checking}.
Span equivalence checking must be performed carefully because even simple bases
in Qwerty may be exponentially large. For example, both
\lstinline!{'0','1'}[64]! and \lstinline!{'1','0'}[64]! both represent $2^{64}$ basis vectors,
meaning a na\"ive algorithm could require $O(2^{64})$ time to type check
\lstinline!{'0','1'}[64] >> {'1','0'}[64]!.
Instead, \compilername{} checks span equivalence in polynomial time.

The key insight in efficient span checking is using factoring --- the opposite
of the na\"ive approach of taking products of lists of vectors.
Factoring appeals to
the definition of tensor product~\cite{nielsen_quantum_2010,axler_linear_2023},
by which
$\texttt{\{}\mathrm{pre}_1\texttt{,}\mathrm{pre}_2\texttt{,}\ldots\texttt{,}\mathrm{pre}_n\texttt{\}}\texttt{ + }\texttt{\{}\mathrm{suff}_1\texttt{,}${\allowbreak}$\mathrm{suff}_2\texttt{,}\ldots\texttt{,}\mathrm{suff}_m\texttt{\}}$ is equal to the following basis: \\
\begin{align*}
\texttt{\{}&\mathrm{pre}_1\texttt{+}\mathrm{suff}_1\texttt{,}\mathrm{pre}_1\texttt{+}\mathrm{suff}_2\texttt{,}\ldots\texttt{,}\mathrm{pre}_1\texttt{+}\mathrm{suff}_m\texttt{,} \\
           &\mathrm{pre}_2\texttt{+}\mathrm{suff}_1\texttt{,}\mathrm{pre}_2\texttt{+}\mathrm{suff}_2\texttt{,}\ldots\texttt{,}\mathrm{pre}_2\texttt{+}\mathrm{suff}_m\texttt{,} \\
           &\ldots\texttt{,} \\
           &\mathrm{pre}_n\texttt{+}\mathrm{suff}_1\texttt{,}\mathrm{pre}_n\texttt{+}\mathrm{suff}_2\texttt{,}\ldots\texttt{,}\mathrm{pre}_n\texttt{+}\mathrm{suff}_m\texttt{\}}
\end{align*}
Factoring recovers the first form given the second form.

\compilername{} performs span
equivalence checking with two queues of basis elements, one queue for each side
of the basis translation. At each iteration, a basis element is popped from each queue, factoring the larger of the two if their dimensions do not match.
After factoring, both basis elements at each iteration must either be identical
or both fully span.
(The check for equality holds because $\varbin$ and $\varbout$ are
initially \textit{normalized} --- vector phases are removed and basis
vectors are sorted lexicographically.)
The efficient span checking algorithm runs in time $O(k^2\log k)$, where $k$ is
the number of AST nodes making up the basis translation. Full details of the
factoring algorithm, including a proof of its time complexity, can be found in Appendix~\ref{app:span-checking}\ofsup{}.

\begin{figure}
    \centering
    \includegraphics[width=0.95\linewidth]{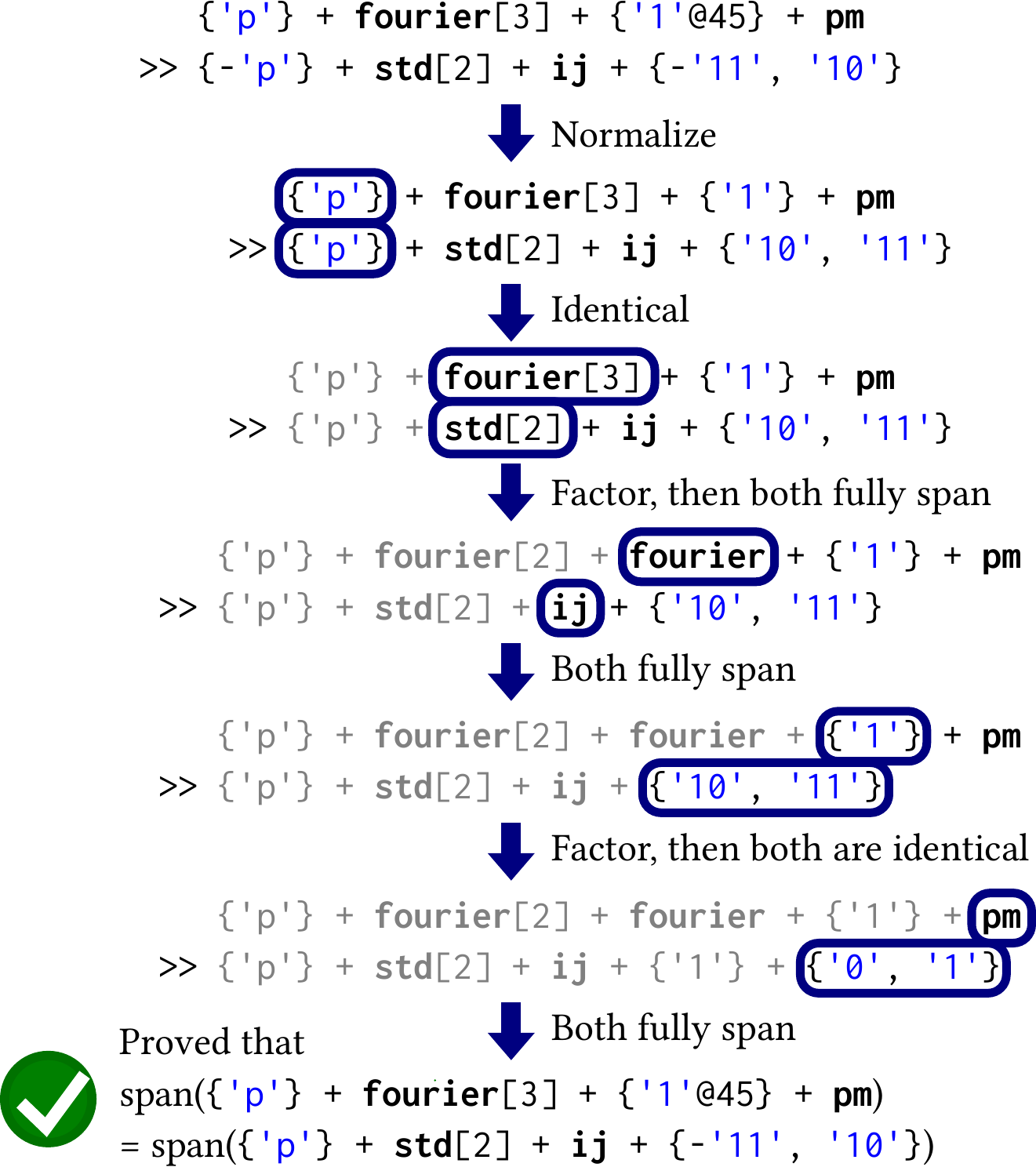}
    \Description{An example of type-checking a basis translation step-by-step.}
    \caption{An example of how \compilername{} type checks a Qwerty basis translation.}
    \label{fig:span-check-example}
\end{figure}

\noindent{\bf Example:}
Fig.~\ref{fig:span-check-example} shows an example of type checking the basis
translation written on the first two lines of the figure.
The blue circles at each step represent the heads of the queues of basis
elements. Note that each rewrite shown does not necessarily represent a
semantically identical basis translation; rather, each step guarantees only
that the span of each side of the translation is unchanged.

\subsection{AST Canonicalization}
After typechecking guarantees that the AST is well-formed, \compilername{}
canonicalizes the AST, performing rewrites such as the following:
\begin{itemize}
\item \textit{\textbf{Removing}} double-adjointing (i.e., replacing \lstinline!~~f! with \lstinline!f!);
\item \textit{\textbf{Rewriting}} \lstinline!std[N] & f! to \lstinline!id[N] + f! (because \lstinline!std[N]! fully spans);
\item \textit{\textbf{Substituting}} \lstinline[morekeywords={b1,b2}]!~(b1 >> b2)! with 
\lstinline[morekeywords={b1,b2}]!b2 >> b1!;  
\item \textit{\textbf{Replacing}} \lstinline[morekeywords={b1,b2,b3}]!b3 & (b1 >> b2)! with \\
\lstinline[morekeywords={b1,b2,b3}]!b3 + b1 >> b3 + b2!; \textit{and}
\item  Performing \textit{\textbf{float constant folding}}. 
\end{itemize}
AST optimizations simplify the IR produced, making the output amenable to
subsequent IR-level optimizations. Additionally, optimizations require less
engineering when done at the AST level --- inside the compiler implementation,
most of these optimizations are ${\sim}5$ lines of code at the AST level versus
${\sim}50$ lines at the MLIR level.

\section{Qwerty IR}\label{sec:qwerty-dialect}
After \compilername{} performs type checking, little other compilation can
happen at the AST level --- a more structured representation is needed. For
instance, while the semantics of Qwerty code can be gleaned from the AST, one cannot
easily reason about dependencies between operations from the AST alone. The
mission of the \texttt{Qwerty} MLIR dialect (or \textit{Qwerty IR}) is to
represent these dependencies while faithfully preserving language semantics.

Qwerty IR is a quantum SSA (static single assignment) IR whose structure is
inspired by prior gate-level IRs such as QIRO and
QSSA~\cite{ittah_qiro_2022,peduri_qssa_2022}. Quantum instructions do not have
side effects; instead, qubits flow through operations (e.g., basis translation
operations), causing dependencies to be explicit in the IR. This naturally
aligns with the linear qubit type in Qwerty~\cite{adams_qwerty_2024} and allows
for optimizations and lowering via simple DAG-to-DAG rewriting.

\noindent{\textbf{Qwerty IR Types:}}
Qwerty IR defines three types: a tuple of \lstinline!N! qubits,
written \lstinline[style=ir]!qbundle[N]!; a tuple of \lstinline!N! bits,
written \lstinline[style=ir]!bitbundle[N]!; and a function type which may be either
reversible or irreversible.
By nature of MLIR allowing mixing operations from multiple dialects, Qwerty IR may
contain types from other dialects, such as the \lstinline[style=ir]!qubit! type from the
QCircuit MLIR dialect (Section~\ref{sec:quantum-types}) or the \lstinline[style=ir]!f64! or
\lstinline[style=ir]!i1! types built into MLIR.

\noindent{\textbf{Qwerty IR Attributes:}}
The most pivotal compile-time constant values (i.e., MLIR attributes) in Qwerty IR are those used to represent the structure of bases:
\texttt{BuiltinBasis}, \texttt{BasisVec\-tor}, \texttt{BasisLiteral},
and \texttt{Basis}. These attributes respectively correspond to the definitions of
built-in basis, basis vector, basis literal, and a canon form of a basis as
defined in Section~\ref{sec:qwerty-intro}.
For example, \lstinline!pm[2] + {'p'}! in Qwerty code corresponds to a \texttt{Basis} with two elements: (1) a \texttt{BuiltinBasis} with primitive basis \lstinline!pm! and dimension 2, and (2) a \texttt{BasisLiteral} containing only one \texttt{BasisVector} with primitive basis \lstinline!pm! and eigenbits \texttt{0}.

\noindent{\textbf{Qwerty IR Operations:}}
The following are the key quantum operations in Qwerty IR:
\begin{itemize}
\item \lstinline[style=ir]!qbprep std<PLUS>[3]!, which prepares a qbundle in the given primitive basis and eigenstate ($\ket{000}$ in this example);
\item \lstinline[style=ir]!qbdiscard %qb!, which resets each qubit in the bundle \texttt{\%qb} and returns the qubits to the ancilla pool;
\item \lstinline[style=ir]!qbdiscardz %qb!, which assumes the qubits are $\ket{0}$ and returns them to the ancilla pool without a reset;
\item \lstinline[style=ir]!qbtrans %qb by b1 >> b2!, which performs the basis translation \lstinline!b1 >> b2! on the qbundle \lstinline!%qb!, returning a new qbundle; and
\item \lstinline[style=ir]!qbmeas %qb in b!, which measures the qbundle \texttt{\%qb} in the basis \texttt{b}.
\end{itemize}
There are also structural operations such as \lstinline[style=ir]!qbpack!, \lstinline[style=ir]!bitpack!, \lstinline[style=ir]!qbunpack!, and \lstinline[style=ir]!bitunpack!, all of which create or destructure qbundles and bitbundles. To support adjointing and predicating functions, respectively, Qwerty IR contains ops such as \lstinline[style=ir]!func_adj %f! and \lstinline[style=ir]!func_pred b %f!, where \texttt{\%f} is a function value and \texttt{b} is a basis. A direct call op may also be marked as an adjoint or predicated call, as in \lstinline[style=ir]!call adj @f()! or \lstinline[style=ir]!call pred(b) @f()!.
There is also an indirect call op that calls a function value, \lstinline[style=ir]!call_indirect %f()!.

\subsection{Lowering a Qwerty AST to Qwerty IR}
With Qwerty IR now defined, we turn to the necessary compilation step of converting a Qwerty AST to Qwerty IR.
Producing Qwerty IR from a Qwerty AST is straightforward overall, but the AST
and IR have some structural differences. First, the AST has a tensor
product node, but Qwerty IR does not have a tensor product op. This is handled
differently depending on the types being tensored together: qbundles are both
\lstinline[style=ir]!qbunpack!ed and then repacked with \lstinline[style=ir]!qbpack! into a combined
qbundle; \texttt{BasisAttr}s have their lists of basis elements concatenated; and
functions are tensored by generating a \lstinline[style=ir]!lambda! op that unpacks the
input qbundle, calls both functions with repacked arguments, unpacks the
result of each, and then returns a repacked combined qbundle.

Another major difference between the AST and IR is that
\discretionary{\lstinline[morekeywords={b1,b2}]!b1 >>!}{\lstinline[morekeywords={b1,b2}]!b2!}{\lstinline[morekeywords={b1,b2}]!b1 >> b2!}
in
Qwerty is a function value, whereas \lstinline[style=ir]!qbtrans! in Qwerty IR is merely an op.
Program semantics are preserved by wrapping the \lstinline[style=ir]!qbtrans! in a
\lstinline[style=ir]!lambda!.
The process for lowering other Qwerty built-in functions like
$b$\lstinline!.measure! for some basis $b$ is similar~\cite{adams_qwerty_2024}. In general, the initial Qwerty IR produced by
the AST walk will never contain \lstinline[style=ir]!call! ops, only \lstinline[style=ir]!call_indirect!
ops, since the Qwerty pipe operator \texttt{|} found in the AST calls function
values, not symbol names.

\subsection{Taking the Adjoint of Basic Blocks}\label{sec:bb-adjoint}
Although Qwerty IR can express a \lstinline[style=ir]!call! \lstinline[style=ir]!adj! \lstinline[style=ir]!@f(%qb)!,
we do not assume hardware has the capability to run code backwards. Instead,
\compilername{} can take the adjoint of a function. This is
done similarly to the Catalyst compiler~\cite{Ittah_Catalyst_a_Python_2024}, except that the algorithm for adjointing a basic block in \compilername{} is not hardcoded for
particular ops. Instead, adjointable ops implement the \texttt{Adjointable} op
interface, which includes a method \texttt{buildAdjoint()} that builds an
adjoint form of a particular op. This way, the Qwerty compiler can traverse the
def-use DAG in a basic block backwards from the block terminator, calling
\texttt{buildAdjoint()} on each op encountered to rebuild a reversed form
top-down.
\begin{figure}
    \centering
    \includegraphics[width=\linewidth]{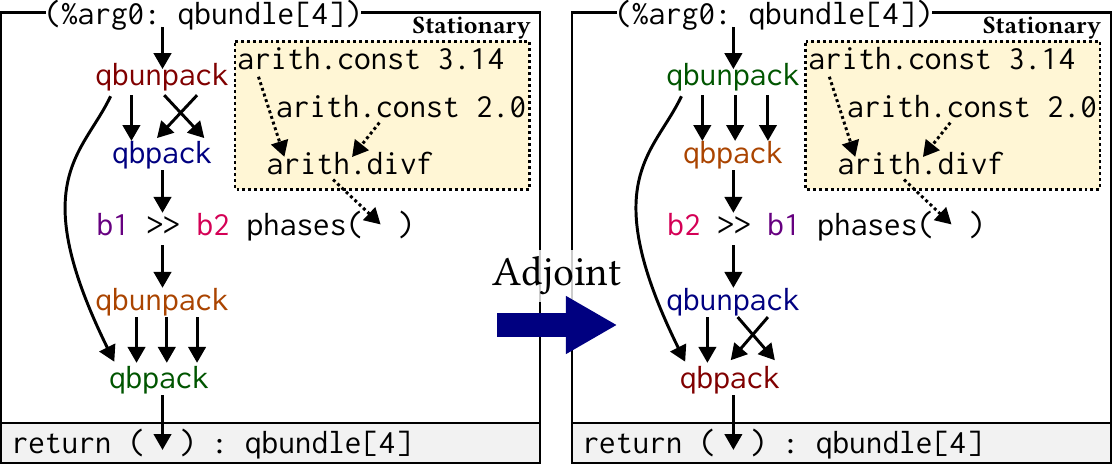}
    \Description{A drawing showing a def-use dag in a basic block reversed.}
    \caption{An example of adjointing a basic block}
    \label{fig:adj-basic-block}
\end{figure}
Classical operations, such as \lstinline[style=ir]!arith.const! ops that define \lstinline[style=ir]!f64!
phase angles, complicate this process because such classical logic may not be
reversible. We call such operations \textit{stationary} because they remain
in-place even if the rest of the DAG (the quantum portion) is inverted around
them. In Fig.~\ref{fig:adj-basic-block}, for example, the classical
\texttt{arith} ops in the yellow box are not adjointed.

\subsection{Predicating Basic Blocks}\label{sec:bb-pred}
\begin{figure}
    \centering
    \includegraphics[width=\linewidth]{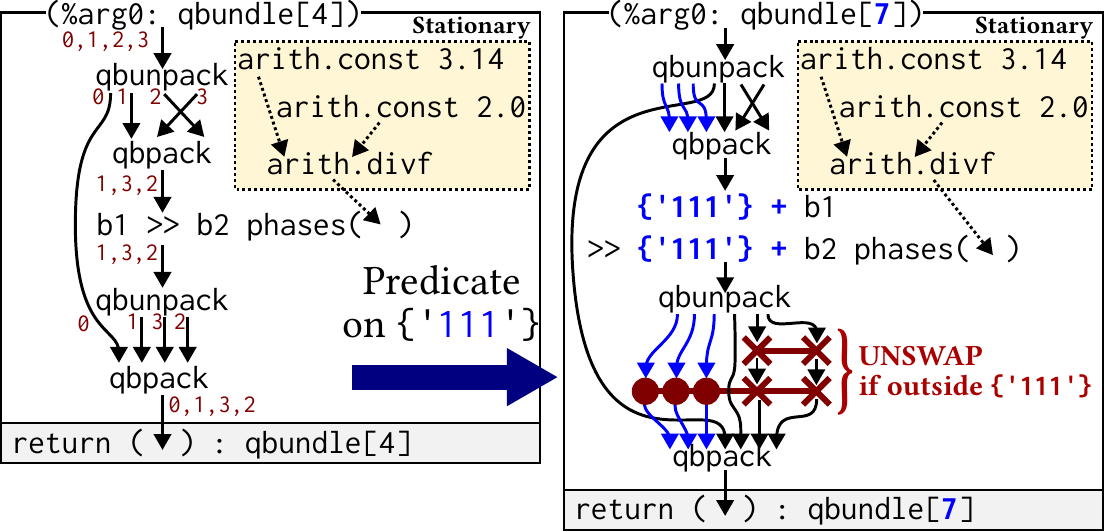}
    \Description{A drawing showing a def-use dag in a basic block predicated on 3 more qubits.}
    \caption{An example of predicating a basic block. The red gates shown on the right are SWAP gates; the second has three controls.}
    \label{fig:pred-basic-block}
\end{figure}

Fortunately, predicating a basic block as needed by a \lstinline[style=ir]!call!
{\lstinline[style=ir]!pred(b) @f(%qb)!}
does not require inverting the
def-use DAG. In most cases, in fact, predication means rebuilding block operations
in-place with new predicates present (e.g., adding the \lstinline!{'111'}!
basis element to both sides of the basis translation in
Fig.~\ref{fig:pred-basic-block}). This is accomplished with a
\texttt{Pre\-di\-cat\-able} op interface that includes a \texttt{buildPredicated()}
method. While adding predicates to all non-stationary ops is necessary, per-op predication is not enough to guarantee the block does not modify states that are orthogonal to the predicate.
This is because the dataflow semantics of Qwerty IR allows effective swaps of
qubits using renaming. In the unpredicated block on the left of
Fig.~\ref{fig:pred-basic-block}, for example, the rightmost two qubits in the
four-dimensional qbundle are swapped by renaming, and this renaming will occur regardless of the new predicates on the subsequent basis translation.

\noindent{\bf Undoing swaps in orthogonal spaces:}
As the final step of predicating a basic block, logical swaps achieved by
renaming must be identified, and cleanup code must be inserted to undo each swap
in the space orthogonal to the predicate.
To accomplish this, \compilername{} runs an intraprocedural dataflow
analysis that maps each MLIR value of type \texttt{qubit}
or \texttt{qbundle}
to a list of
qubit indices that the value represents. The unpredicated block in
Fig.~\ref{fig:pred-basic-block} shows an example of these indices drawn in red
on each non-stationary value edge. The analysis begins by assigning fresh
indices to each non-stationary argument to the basic block.
These indices are then propagated across the
def-use chains in the basic block.
When the analysis completes, the permutation effected by the unpredicated block
is decomposed into a series of swaps. This sequence of swaps is
reversed so that the permutation is undone, and then two SWAP gates are emitted
for each: one uncontrolled SWAP to unconditionally undo the logical swap across
the whole state space, and a second predicated SWAP to redo the swap inside the
predicated space. (This is a trick to undo the original renaming-based swap inside the space orthogonal
to the predicate.)

An example of this swap-unswap gate sequence is shown in red in the
bottom right of Fig.~\ref{fig:pred-basic-block}. 
Observe that when the leftmost three qubits are $\ket{111}$, the two red SWAP gates in Fig.~\ref{fig:pred-basic-block} will cancel out, leaving the rightmost two qubits swapped by renaming as desired.
If the predicate qubits are in a different computational basis state such as $\ket{101}$ (i.e., the function runs on a state orthogonal to the predicate), the second (controlled) SWAP gate will not execute, making the first SWAP gate cancel out the initial swap achieved by renaming.

\subsection{Qwerty IR Optimizations}\label{sec:inlining}
To target the broadest set of quantum hardware, it is desirable to convert
functional Qwerty code into a straight-line sequence of quantum operations if
possible. To this end, inlining is the most important optimization in the
Qwerty compiler.
However, despite \lstinline[style=ir]!call_indirect! ops  mirroring Qwerty
semantics, indirect calls are more clumsy to inline, so it is desirable to replace as many
\lstinline[style=ir]!call_indirect!s as possible with direct \lstinline[style=ir]!call!s instead.
This can be challenging in the presence of \lstinline[style=ir]!lambda!s, which cannot be directly converted to \lstinline[style=ir]!call!s because they have no
symbol name to reference.

The compiler overcomes the brunt of these issues with the following
sequence of passes: (1) lifting all \lstinline[style=ir]!lambda!s to \lstinline[style=ir]!func!s
referenced by \lstinline[style=ir]!func_const!s; (2) running the MLIR canonicalizer, which
we configure to convert every instance of
\discretionary{\lstinline[style=ir,morekeywords={func_}]!call_indirect(func_!}{\lstinline[style=ir,morekeywords={const}]!const @f)()!}
              {\lstinline[style=ir]!call_indirect(func_const @f)()!}
into \lstinline[style=ir]!call @f()!;
and (3) running the MLIR inliner, which repeatedly inlines and re-runs the
canonicalizer to find new inlining opportunities. This canonicalization
includes patterns in the IR involving \lstinline[style=ir]!func_adj! or \lstinline[style=ir]!func_pred!,
such as
\discretionary{\lstinline[style=ir]!call_indirect(!}{\lstinline[style=ir]!func_pred {'10'} (func_adj (func_const @f)))()}!}
              {\lstinline[style=ir]!call_indirect(func_pred {'10'} (func_adj (func_const @f)))()}!}, \\
which becomes \lstinline[style=ir]!call adj pred({'10'}) @f()!.
When calls with \lstinline[style=ir]!adj! or \lstinline[style=ir]!pred! are inlined, the routines
in Sections~\ref{sec:bb-adjoint} and \ref{sec:bb-pred} are used to transform
the single basic block making up the callee function body.
Despite the general effectiveness of this approach, some Qwerty code can still
produce IR that is tricky to inline, which we mitigate with specialized canonicalization patterns. One such inlining-enabling pattern pushes \lstinline[style=ir]!call_indirect! ops whose callee is defined by an \lstinline[style=ir]!scf.if! op upward into both forks of the \lstinline[style=ir]!scf.if! op (see Appendix~\ref{app:inlining-cond}\ofsup{} for details).

\section{QCircuit Dataflow IR}\label{sec:quantum-dialect}
While the high-level semantics of Qwerty IR are valuable for analysis or
transformations at that abstract level, directly lowering Qwerty IR to an output
format such as QIR would be quite tedious and likely suboptimal.
\compilername{} fills the gap between Qwerty IR and an optimized output circuit
using another level of IR: the \texttt{QCircuit} MLIR dialect (or
\textit{QCircuit IR}), a gate-level dataflow-semantics
quantum IR similar to QIRO or QSSA \cite{ittah_qiro_2022,peduri_qssa_2022}. It
takes particular inspiration from McCaskey and Nguyen's MLIR
dialect~\cite{mccaskey_mlir_2021}, which also acts as a sort of MLIR dialect
for QIR.

\noindent{\bf QCircuit IR Types:}\label{sec:quantum-types}
QCircuit IR has three types: \lstinline[style=ir]!qubit!,
\lstinline[style=ir]!array<!$T$\lstinline[style=ir]!>[N]!, and
\lstinline[style=ir]!callable!. These types correspond to the \texttt{\%Qubit*},
\texttt{\%Array*}, and \texttt{\%Callable*} types in QIR.

\noindent{\bf QCircuit IR Operations:}\label{sec:quantum-dialect-ops}
The core operations in QCircuit IR are the following:
\begin{itemize}
\item \lstinline[style=ir]!qalloc!, which allocates a qubit;
\item \lstinline[style=ir]!qfree %q!, which frees a qubit;
\item \lstinline[style=ir]!qfreez %q!, which is similar to \lstinline[style=ir]!qfree! but assumes the qubit is $\ket{0}$ and skips the reset performed by \lstinline[style=ir]!qfree!;
\item \lstinline[style=ir]!measure %q!, which measures a qubit in the standard basis, yielding the new state and an \lstinline[style=ir]!i1! result; and
\item \lstinline[style=ir]!gate G [%c1,...,%cM] %q1,...,%qN!, which runs a controlled \texttt{G} gate, taking $\texttt{M}+\texttt{N}$ qubits and yielding their new states. The operands in square brackets $\texttt{[}\cdot\texttt{]}$ are controls and the rest are targets.
\end{itemize}
The structural \lstinline[style=ir]!arrpack! and \lstinline[style=ir]!arrunpack! ops facilitate code that manipulates arrays, being analogous
to the \lstinline[style=ir]!qbpack! and \lstinline[style=ir]!qbunpack! ops for qbundles in Qwerty IR. To support callable values, QCircuit IR includes ops for holding callable value metadata and for invoking, controlling, or adjointing \lstinline[style=ir]!callable!s; these correspond closely to both the function ops in Qwerty IR and to QIR callable intrinsics~\cite{qir_callables}.
Thanks to canonicalization and inlining (Section~\ref{sec:inlining}),
ops for arrays and callables are rarely used in realistic QCircuit-dialect IR
generated by the \compilername{} compiler. To our knowledge, though,
\compilername{} is the first MLIR-based compiler to generate QIR
callables~\cite{qir_callables}.

\subsection{Lowering Qwerty IR to QCircuit IR}
Conversion from Qwerty IR to QCircuit IR is performed using the MLIR
dialect conversion framework~\cite{lattner_mlir_2021}, in which conversions are defined using
rewrite patterns. For example, we define a conversion pattern to
decompose the Qwerty \lstinline[style=ir]!qbdiscard! op into a \lstinline[style=ir]!qbunpack! that feeds
into $N$ QCircuit-dialect \lstinline[style=ir]!qfree! ops. (The Qwerty \lstinline[style=ir]!qbunpack! is then
converted into a QCircuit \lstinline[style=ir]!arrunpack!.) Lowering \lstinline[style=ir]!qbprep! ops is
similarly simple, with them being broken into $N$ \lstinline[style=ir]!qalloc!s followed by
\lstinline[style=ir]!gate! ops for $H$, $S$, and $X$ gates, all of whose
results are \lstinline[style=ir]!qbpack!ed into the substituted result. A canonicalization
pattern later removes redundant instances of
\lstinline[style=ir]!arrunpack(!{\allowbreak}\lstinline[style=ir]!arrpack)! and \lstinline[style=ir]!arrpack(arrunpack)!,
usually leaving a clean DAG starting with \lstinline[style=ir]!qalloc!s and ending in
\lstinline[style=ir]!qfree!s.
Lowering Qwerty IR in its full generality only becomes challenging when
inlining fails or if programmers write nontrivial basis translations. We cover
these issues in Sections~\ref{sec:func-spec-gen} and \ref{sec:lower-btrans}.
Section~\ref{sec:classical-synth} then describes how \compilername{} uses
Tweedledum to synthesize QCircuit dialect code from classical circuit
descriptions~\cite{schmitt_tweedledum_2022}. Section~\ref{sec:quantum-opt} concludes by describing
optimizations we perform in the QCircuit dialect.

\subsection{Producing Function Specializations}\label{sec:func-spec-gen}
One challenge in lowering Qwerty IR to QCircuit IR is that a function value in
Qwerty IR cannot be represented by a typical function pointer --- i.e., a mere address of code in memory --- in QCircuit IR. This is because a function value \texttt{\%f}
in Qwerty IR can be adjointed with \lstinline[style=ir]!func_adj %f! or predicated on a basis
\texttt{b} with \lstinline[style=ir]!func_pred b %f!, each
producing a new function value with fundamentally different structure. Thus,
the compiler may need to generate multiple \textit{specializations} of a given
function to satisfy these requests~\cite{qsharp_specializations}.

\noindent{\textbf{Function specialization analysis:}}
The most na\"ive conservative generation of all specializations for all
functions is intractable because there are an enormous number of
possible predicated specializations given different predicates.
To avoid this, we run interprocedural dataflow analysis \cite{mlir_dataflow} on
the IR to to label each callable value with a set of tuples
$(\mathrm{funcName},\mathrm{isAdjoint},\mathrm{numControls})$.
Then, the analysis for the callee operand can be checked at every
\lstinline[style=ir,morekeywords={call_}]!call_!{\allowbreak}\lstinline[style=ir,morekeywords={indirect}]!indirect! op, determining (along with every \lstinline[style=ir]!call! op) the specializations of other functions that a forward (non-\lstinline[style=ir]!adj! and non-\lstinline[style=ir]!pred!) invocation of each function requires.

\noindent{\bf Transitive specialized calls:}
However, this analysis does not account for all paths through the call graph;
specifically, by considering the callees of only a forward invocation of each
function, it does not consider transitive specialization requirements. Consider
the following IR, for instance:
\begin{lstlisting}[style=irxs]
func f() {           func g() {        func h() {
   call adj g(...)      call h(...)       ...
}                    }                 }
\end{lstlisting}
An adjoint specialization of \texttt{h()} is needed because the adjoint form of
\texttt{g()} is called by \texttt{f()}. However, this would not be detected by
the initial analysis because there is no explicit \lstinline[style=ir]!call adj h()! in the
IR. \compilername{} addresses this problem by constructing a call graph of
specializations and iteratively addding transitive edges as required. (See
Appendix~\ref{app:specgen}\ofsup{} for the full algorithm.)

\subsection{Lowering Basis Translations to Gates}\label{sec:lower-btrans}
The toughest challenge in lowering Qwerty IR to QCircuit IR is synthesizing the
quantum gates that achieve a basis translation. This is the most novel part of
\compilername{}, in fact, as we are not aware of prior compilers that
synthesize basis translations, only implementations that do more general (and
more expensive) unitary synthesis~\cite{shende_synthesis_2006}.

\begin{figure}
\includegraphics[width=\linewidth]{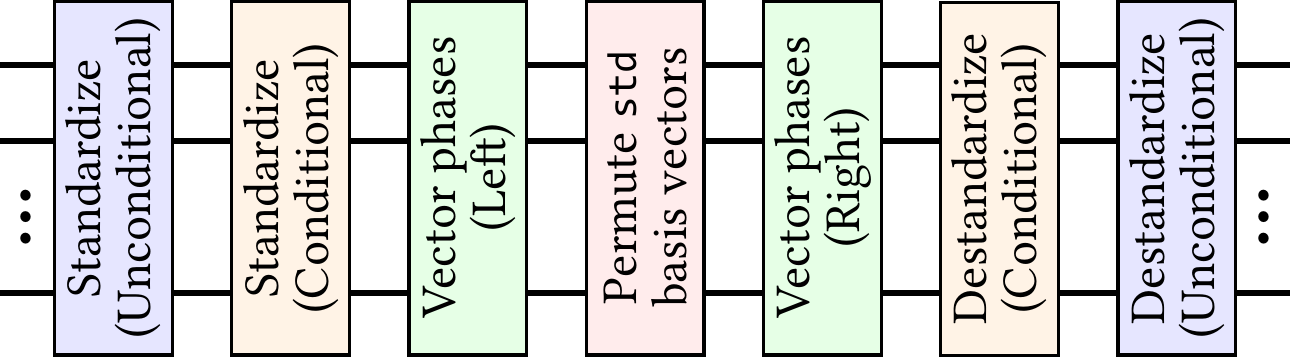}
\Description{A figure showing the structure of a synthesized basis translation: (de)standardization on the outside, then vector phases inside, and then a permutation of std vectors inside of that}
\caption{A summary of the structure of the circuit synthesized by \compilername{} for a basis translation (read left to right)
}\label{fig:synth-high-level}
\end{figure}

As seen in Fig.~\ref{fig:synth-high-level}, the core of a basis translation is a
permutation of \lstinline!std! basis vectors. Yet because the left-hand side of
a basis translation can be defined in terms of non-\lstinline!std! primitive
bases, standardization runs first to translate qubits from that primitive basis
to \lstinline!std!. (Destandardization, which runs last, translates from
\lstinline!std! to the primitive basis on the right-hand side of the goal
translation.) The phases that programmers supply for basis vectors are handled
between the steps of (de)standardizing and permutation since a typical gateset
makes imparting phases most convenient in the standard basis.
The following sections describe the steps in synthesis.

\noindent{\textbf{Standardization:}}%\label{sec:stdize}
\begin{figure}
\begin{adjustbox}{width=0.87\linewidth}\begin{minipage}{\linewidth}\centering\input{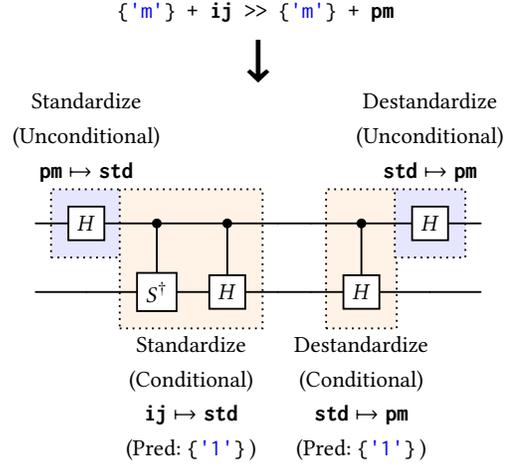}\end{minipage}\end{adjustbox}
\Description{A figure showing an example of conditional standardization}
\caption{An example of basis translation circuit synthesis by \compilername{} involving both conditional and unconditional standardization}\label{fig:cond-stdize-example}
\end{figure}
Performing standardization for a basis translation $\varbin$\texttt{ >{}> }$\varbout$ begins by
collecting two lists of pairs $(\mathrm{prim},\mathrm{dim})$, where $\mathrm{prim}$ is a primitive basis and $\mathrm{dim}$ is the number of qubits, one list for $\varbin$ and the other for $\varbout$.
These lists represent the ranges of primitive bases across basis elements of
$\varbin$ and $\varbout$, respectively, and the first is a list of \textit{necessary standardizations} while the second is a list of \textit{necessary destandardizations}. The former list specifies how primitive bases in $\varbin$ must be translated to \lstinline!std! for all operand qubits at the beginning of synthesis, and the latter list specifies how \lstinline!std! must be translated to the primitive bases in $\varbout$ at the end of synthesis.
Each necessary (de)standardization is annotated as either \textit{unconditional} or
\textit{conditional}, indicating whether it is present in both the list of
necessary standardizations and destandardizations, i.e., both sides of the goal
translation. Unconditional standardizations represent an unchanged primitive
basis on either side of the translation, whereas conditional
(de)standardizations represent a change in primitive basis.

The distinction between conditional and unconditional (de)stan\-dard\-ization
matters when a basis translation has predicates. The gates that perform unconditional standardization do not
need to be controlled such that they run only in the space on which the goal
translation is predicated, since unconditional standardizations will be undone
by their corresponding destandardizations.
(In quantum parlance, the rest of the synthesized circuit is \textit{conjugated} by unconditional (de)standardizations.)
For instance, if the upper qubit in
Fig.~\ref{fig:cond-stdize-example} is initially $\ket{+}$, the controlled gates
shown will not run and the $H$ gates on the upper qubit will cancel out, leaving
the state correctly unaffected.
On the other hand, the controlled gates in
Fig.~\ref{fig:cond-stdize-example}, which achieve a conditional standardization, could not be replaced with uncontrolled gates because the circuit would incorrectly modify the input state if the
upper qubit is e.g. $\ket{+} \not\in \vspan(\ket{-})$. (In general, the erroneous
circuit would behave as\;
\discretionary{\lstinline!{'p','m'} + ij >> {'p',!}{\lstinline!'m'} + pm!}
              {\lstinline!{'p','m'} + ij >> {'p','m'} + pm!}
instead of the goal translation shown in Fig.~\ref{fig:cond-stdize-example}.)
The full process of determining which (de)standardizations are conditional is shown in
Algorithm~\ref{alg:standardize} in Appendix~\ref{app:standardize}\ofsup{}. 

Once (de)standardizations and their conditionality are determined, im\-ple\-menting un\-conditional
standardizations is straightforward using $H$ and $S$ gates (for
\lstinline!pm! and \lstinline!ij!) or $N$-bit IQFT (inverse quantum Fourier
transform, for \lstinline!fourier[!$N$\lstinline!]!).
Unconditional destandardization is simply
the adjoint of these operations.
Synthesizing conditional (de)standardiza\-tions requires controlled forms of
these, yet in typical gatesets, controlling on $\ket{0}$ and $\ket{1}$ is more
convenient than controlling on vectors in other primitive bases. Fortunately,
span equivalence checking (Section~\ref{sec:span-equiv-checking}) guarantees
that all predicates in a well-typed basis translation correspond to
unconditional standardizations.
Thus, \compilername{} synthesizes an outer layer of unconditional
(de)standardizations before synthesizing an inner layer of controlled
conditional (de)standardiza\-tions, as seen in
Fig.~\ref{fig:cond-stdize-example}.

\noindent{\bf Vector Phases:}%\label{sec:vec-phase}
\begin{figure}
\begin{adjustbox}{width=0.87\linewidth}\begin{minipage}{\linewidth}\centering\input{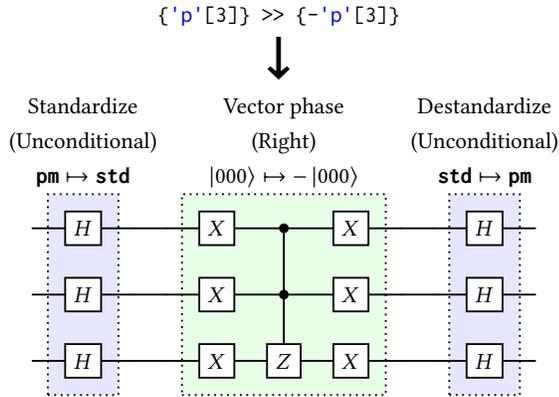}\end{minipage}\end{adjustbox}
\Description{A figure showing an example of a vector phase}
\caption{An example of basis translation circuit synthesis involving an vector phase}\label{fig:vec-phase-example}
\end{figure}
Given the goal basis translation $\varbin$\texttt{ >{}> }$\varbout$, the
compiler scans each of $\varbin$ and $\varbout$ for basis vectors with phases.
The result is two lists of tup\-les $(\mathrm{eigenbits}, \theta,
\mathrm{offset})$, one list for the input basis $\varbin$ and one for the
output basis $\varbout$. As shown in Fig.~\ref{fig:synth-high-level}, after
standardization, the input state is translated from \lstinline!std! with the
input phases to \lstinline!std! without phases. Then any desired phases are
re-introduced before destandardization via translation from \lstinline!std! to
\lstinline!std! with the output phases.
These translations are achieved with $X$-conjugated
multi-controlled--$P(\theta)$ gates.
Fig.~\ref{fig:vec-phase-example} shows
an example of synthesizing a translation that uses vector phases to
implement a three-qubit Grover
diffuser~\cite{grover_fast_1996,grover_quantum_1997,nielsen_quantum_2010}.

\noindent{\bf Permutation:}%\label{sec:synth-perm}
\begin{figure}
\begin{adjustbox}{width=0.87\linewidth}\begin{minipage}{\linewidth}\centering\input{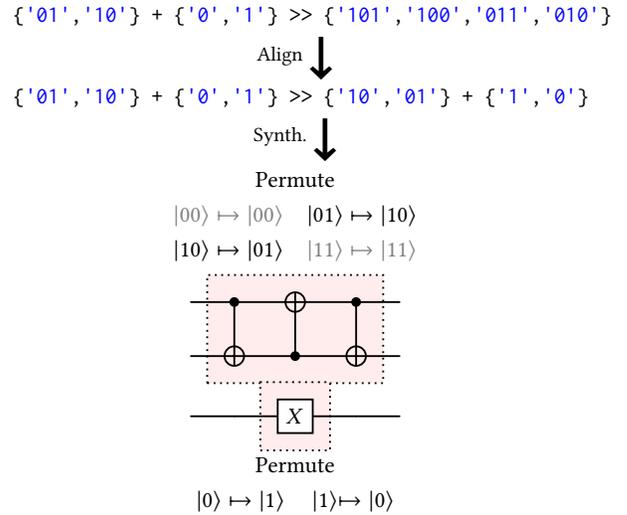}\end{minipage}\end{adjustbox}
\Description{A figure showing an example of a simple permutation}
\caption{An example of basis translation circuit synthesis involving a permutation}\label{fig:perm-example}
\end{figure}
With standardization and vector phases handled, only the center of
Fig.~\ref{fig:synth-high-level} remains: synthesizing a circuit that implements
the reversible classical function $f : \mathbb{B}^n \rightarrow \mathbb{B}^n$,
which acts as a permutation on \lstinline!std[!$n$\lstinline!]! basis vectors.
Unless the permutation is the identity,
this synthesis is performed for each pair of respective basis elements in the
two bases. (If respective bases elements are different dimension and thus do
not operate on the same qubits, a factoring-based process of \textit{alignment}
is necessary. This is discussed in detail in Appendix~\ref{app:align}\ofsup{}.) In
Fig.~\ref{fig:perm-example}, for instance, synthesis runs twice: once for the
permutation defined for the first two qubits
and again for the permutation defined for the last qubit.
Presently, \compilername{} uses the multidirectional transformation-based
synthesis algorithm~\cite{miller_transformation_2003,soeken_fast_2016}
implemented in the Tweedledum library~\cite{schmitt_tweedledum_2022}.
Controls are added to the Tweedledum-generated circuit as necessary to obey the
predicates in aligned bases.

\subsection{Synthesizing Classical Functions}\label{sec:classical-synth}
Basis-oriented constructs are not the only way that Qwerty programmers can
express program code, since Qwerty also supports synthesizing circuits from
classical logic~\cite{adams_qwerty_2024}.
The \lstinline!@classical! function on lines~\lnbvsimpleclassical{}-\lnbvsimpleclassicalret{} of
Fig.~\ref{fig:bv_simple} is an example of this feature.
\compilername{} begins compiling \lstinline!@classical! functions similarly as it does for \lstinline!@qpu!
functions (Section~\ref{sec:ast}), by extracting a custom AST from the Python AST and performing type checking~\cite{ast}.
This AST is then converted to a classical logic network in the mockturtle library~\cite{soeken_epfl_2022}. After
running some classical optimizations
provided by mockturtle, \compilername{} passes this classical circuit to
tweedledum~\cite{schmitt_tweedledum_2022}, which generates a Bennett embedding
$U_f\ket{x}\ket{y} = \ket{x}\ket{y \oplus f(x)}$ for the corresponding
classical function $f$
\cite{bennett_timespace_1989,meuli_role_2019,schmitt_compilation_2021}. The
result is expressed in tweedledum \texttt{Circuit} IR, which we transpile to
QCircuit IR.
Inside a Qwerty \lstinline!@qpu! function, the programmer can request this Bennett
embedding for \lstinline!f! with the syntax \lstinline!f.xor!. The syntax
\lstinline!f.sign! generates the form $U_f'\ket{x} = (-1)^{f(x)}\ket{x}$
instead, which \compilername{} generates by passing $\ket{-}$ ancilla to
the output of the Bennett embedding~\cite{nielsen_quantum_2010}.

\subsection{QCircuit IR Optimizations}\label{sec:quantum-opt}
\begin{figure}
\includegraphics[width=\linewidth]{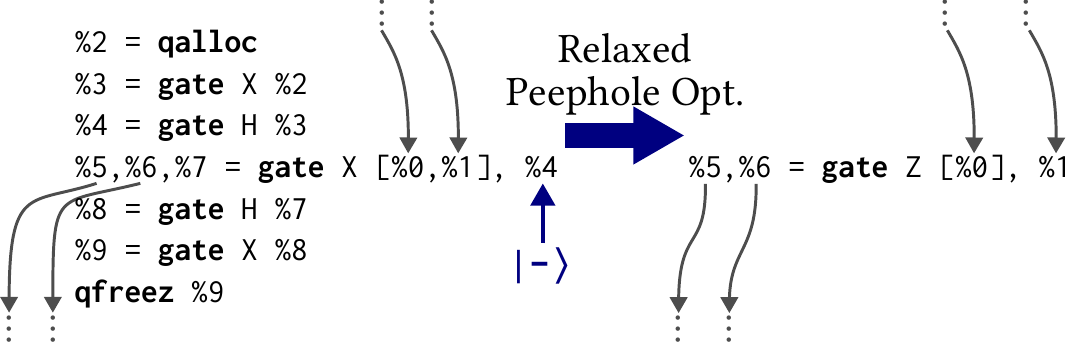}
\Description{IR in which a multi-controlled--X targeting a $\ket{-}$ state is replaced with a multi-controlled-Z gate}
\caption{A relaxed peephole optimization performed on QCircuit-dialect IR}\label{fig:relaxed-example}
\end{figure}

Before generating the output circuit from QCircuit IR, we run optimizations to
remove redundancies left behind by earlier systematic lowering. For instance,
in Fig.~\ref{fig:cond-stdize-example}, the adjacent controlled-Hadamards
generated by basis translation circuit synthesis should cancel out. We
implement similar common gate-level optimizations as those in prior
work~\cite{ittah_qiro_2022,peduri_qssa_2022}, such as cancelling out adjacent
Hermitian (self-adjoint) gates or replacing $HXH$ with $Z$. We also
implement some relaxed peephole optimizations due to Liu, Bello, and
Zhou~\cite{liu_relaxed_2021}, including the one shown in
Fig.~\ref{fig:relaxed-example}, which turns a multi-controlled $X$ with a $\ket{-}$ target into a multi-controlled $Z$ without the $\ket{-}$ ancilla.
This particular optimization is especially useful for simplifying instances of
\lstinline!f.sign! (used in Bernstein--Vazirani or Grover's), which typically
involves the situation shown in Fig.~\ref{fig:relaxed-example} (see
Section~\ref{sec:classical-synth}).
As a final optimization step, multi-controlled gates are decomposed using
Selinger's controlled-$iX$ scheme to reduce $T$ gate counts on fault-tolerant
hardware~\cite{selinger_quantum_2013}.

\section{Targeting OpenQASM 3 and QIR}\label{sec:targeting}
\noindent\textbf{Transpiling to OpenQASM 3:} From QCircuit IR, \compilername{} can produce OpenQASM~3~\cite{cross_openqasm_2022} using a
process akin to \texttt{reg2mem} in QSSA~\cite{peduri_qssa_2022}, in which SSA
(static single assignment) values are converted to quantum register accesses.
Because OpenQASM 3 does not support function pointers or qubit allocation
inside subroutines, OpenQASM 3 generation is currently dependent on inlining
succeeding (Section~\ref{sec:inlining}), although this has not empirically presented a problem for realistic Qwerty programs (see Section~\ref{sec:eval-inline}).

\noindent\textbf{Producing QIR:} \compilername{} produces QIR by using MLIR dialect conversion from the
QCircuit MLIR dialect to the built-in MLIR dialect named \texttt{llvm}, which
models LLVM IR~\cite{qir, lattner_llvm_2004}. For the Unrestricted QIR profile (in which the full
generality of QIR is allowed), this means individually converting each
QCircuit op into calls to the corresponding QIR intrinsics. For example,
\lstinline[style=ir]!qalloc! becomes a call to
\texttt{\_\_quantum\_\_rt\_\_qubit \\ \_allocate()}, and a \lstinline[style=ir]!gate! is converted to a
call to the corresponding gate intrinsic.
Callable-related ops are lowered to calls to equivalent callable intrinsics. \compilername{}-generated QIR requires only one
change to the QIR runtime that amounts to removing approximately 50 lines of
Q\#-specific runtime code (see Appendix~\ref{app:qir-changes}\ofsup{} for details).

\noindent{\bf Targeting the QIR base profile:}
In contrast with the Unrestricted QIR profile, the QIR Base Profile is highly
restricted~\cite{qir}, effectively amounting to a straight-line
sequence of gates embedded in LLVM IR. In the Base Profile, there is no dynamic
qubit allocation; instead, \texttt{\%Qubit*}s passed to gate intrinsics are
integral qubit indices casted to pointers. Thus, a similar \texttt{reg2mem}
process as used in OpenQASM~3 export is
used to insert the LLVM \texttt{inttoptr} casts that act as \lstinline[style=ir]!qalloc!s~\cite{peduri_qssa_2022}.

\noindent{\bf Execution:}
The Qwerty runtime included with \compilername{} is preconfigured to target either local simulation with
qir-run\-ner \cite{qir_runner} or a va\-ri\-ety of backends via
QIR-EE~\cite{wong_cross-platform_2024}, a convenient adaptor for executing QIR via XACC~\cite{mccaskey_xacc_2020}. \compilername{} produces Unrestricted
Profile QIR for qir-runner but Base Profile QIR for QIR-EE.
(\compilername{}-generated QIR could also be sent to cloud services such as Azure Quantum provided that LLVM versions sufficiently align.)

\section{Evaluation}\label{sec:eval}
\begin{figure*}
\centering
\subcaptionbox{Bernstein--Vazirani\label{fig:eval-runtime:bv}}
{\includegraphics[width=.24\linewidth]{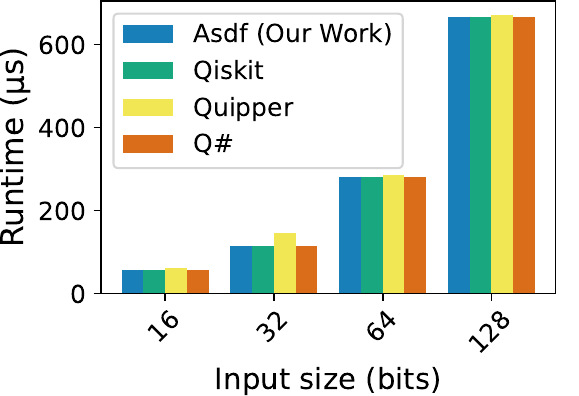}}
\subcaptionbox{Grover's\label{fig:eval-runtime:grover}}
{\includegraphics[width=.24\linewidth]{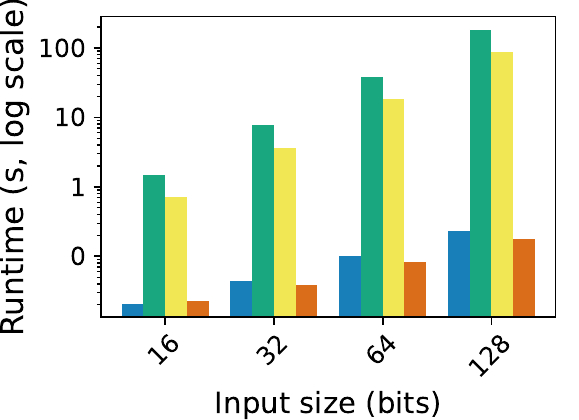}}
\subcaptionbox{Simon's\label{fig:eval-runtime:simon}}
{\includegraphics[width=.24\linewidth]{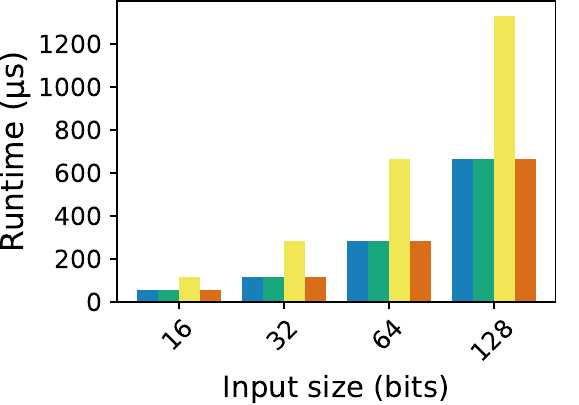}}
\subcaptionbox{Period finding\label{fig:eval-runtime:period}}
{\includegraphics[width=.24\linewidth]{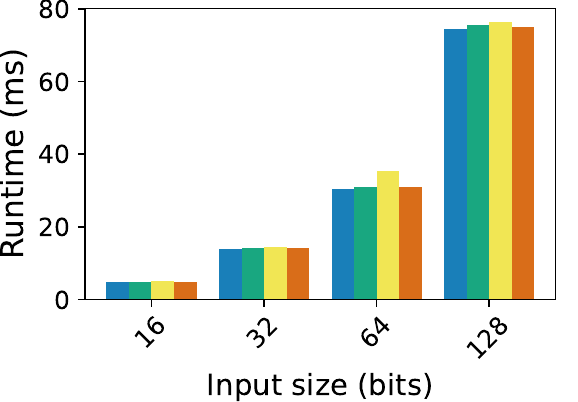}}
\caption{Runtime of benchmarks for each compiler for different oracle input sizes (lower is better)}\label{fig:eval-runtime}
\end{figure*}
\begin{figure*}
\centering
\subcaptionbox{Bernstein--Vazirani\label{fig:eval-qubits:bv}}
{\includegraphics[width=.24\linewidth]{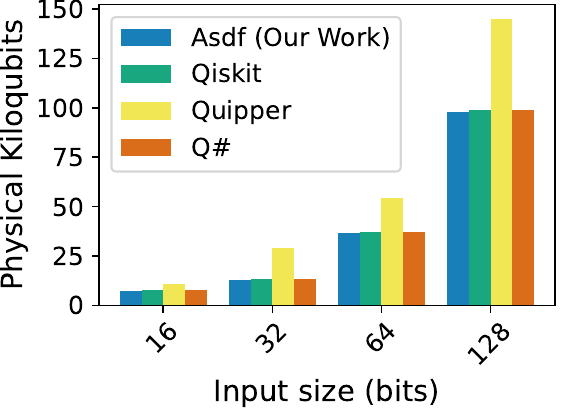}}
\subcaptionbox{Grover's\label{fig:eval-qubits:grover}}
{\includegraphics[width=.24\linewidth]{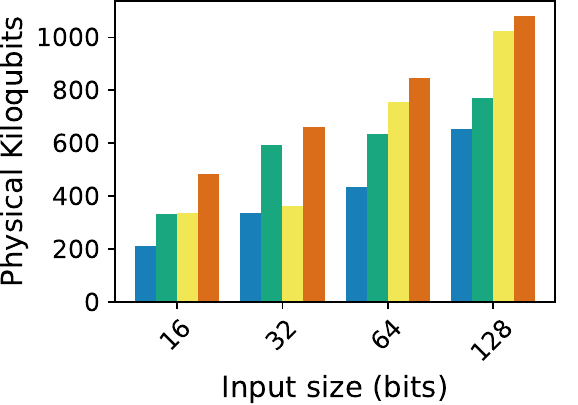}}
\subcaptionbox{Simon's\label{fig:eval-qubits:simon}}
{\includegraphics[width=.24\linewidth]{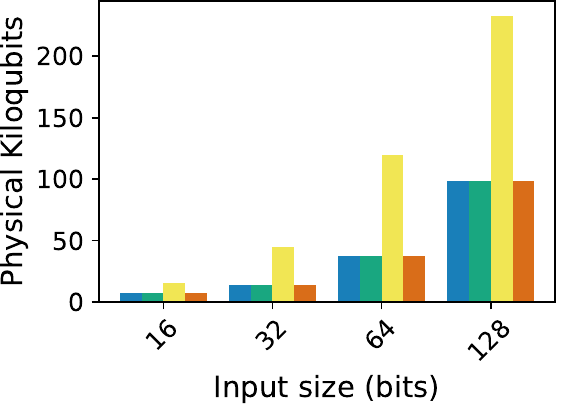}}
\subcaptionbox{Period finding\label{fig:eval-qubits:period}}
{\includegraphics[width=.24\linewidth]{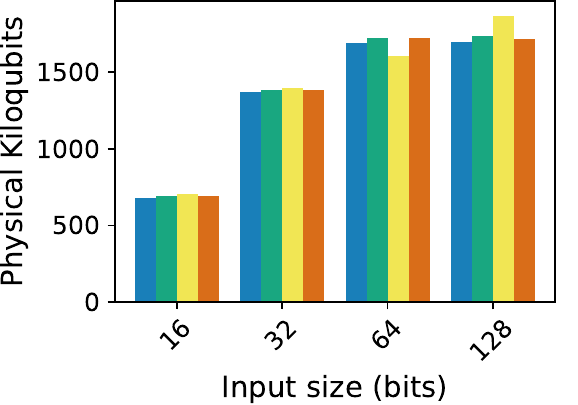}}
\caption{Physical qubits used by each benchmark by each compiler for different oracle input sizes (lower is better)
}\label{fig:eval-qubits}
\end{figure*}

To our knowledge, no other compilers for basis-oriented quantum
programming languages exist, including Qwerty, so we evaluate \compilername{} by
comparing it with compilers for circuit-oriented languages. First, in
Section~\ref{sec:eval-inline}, we investigate the effectiveness of inlining in
\compilername{} (Section~\ref{sec:eval-inline}) by comparing with a Q\#
compiler. Then, in Section~\ref{sec:eval-gates}, we compare the cost of Qwerty
code versus circuit-oriented code in terms of metrics for fault-tolerant
quantum hardware. (Note that the intention of \compilername{} is to
synthesize circuits with similar performance to handwritten circuits,
not to outperform handwritten circuits.)

\subsection{Experimental Setup}\label{sec:eval-setup}
\noindent\textbf{Software packages used:}
We evaluate using Python 3.11.2, Qiskit 1.2.0, Quipper 0.9.0.0, Classic Q\# QDK
0.27, Modern Q\# QDK 1.6, LLVM 19.1.2, and Haskell 8.6.5. The
\texttt{quipper-qasm} tool was used to convert Quipper code to OpenQASM~\cite{quipper-qasm}. We
modified the Qiskit \texttt{-O3} transpiler to remove a few particular passes
that were causing nondeterminism in evaluation results. Before removal, these
passes had a maximum of ${<}0.1\%$ positive performance impact, so the impact
on our evaluation is negligible.

\noindent\textbf{Benchmarks:}
We test with five popular quantum algorithms: Bern\-stein--Vazirani with an
alternating secret string
($1010\cdots$)~\cite{bernstein_quantum_1993,fallek_transport_2016},
Deutsch-Jozsa with a balanced oracle that XORs all input bits
\cite{deutsch_rapid_1997,nielsen_quantum_2010}, Grover's search with an oracle
matching the input consisting of all
1s~\cite{grover_fast_1996,grover_quantum_1997,nielsen_quantum_2010}, Simon's
algorithm with a nonzero secret
string~\cite{simon_power_1994,simon_power_1997}, and QFT-based period finding
with an oracle performing bitmasking~\cite{nielsen_quantum_2010}. All benchmarks are parameterized on the size of a classical input to their oracle.
To make the
evaluation feasible, the Grover's benchmark is capped at 12 iterations. For all
benchmarks, oracles are expressed as classical logic in both Quipper and
Qwerty, but as gates in Qiskit and Q\#. The Q\# code is largely taken
from Wojcieszyn~\cite{wojcieszyn_introduction_2022}; the Qiskit implementations
are derived from the original Qiskit textbook~\cite{wootton_teaching_2021}; the Quipper
implementations are inspired by Siddiqui et al.~\cite{siddiqui_five_2014} and
Green et al.~\cite{green_introduction_2013}; and the Qwerty programs are
derived from the examples by Adams et al.~\cite{adams_qwerty_2024}.

\subsection{Impact of Inlining}\label{sec:eval-inline}
If a compiler targets QIR, a good measure of its success in inlining is how
many QIR callable intrinsics it emits. Presently, the QIR callables API is not
well supported by hardware partners, so fewer calls to callables intrinsics is
preferred. We compare \compilername{} with inlining enabled
(Section~\ref{sec:inlining}), \compilername{} with inlining disabled, and the
Classic Q\# QDK. (We use the Classic Q\# QDK rather than the Modern Q\# QDK because
at the time of writing, the Modern QDK cannot yet generate QIR callables.)
Table~\ref{tab:callables} shows the results. The metrics shown on the
``create'' and ``inv.'' columns of Table~\ref{tab:callables} are the number
of invocations of \texttt{\_\_quantum\_\_rt\_\_callable\_create} and
\texttt{\_\_quantum\_\_rt\_\_callable\_invoke}, respectively, in the \\ LLVM
assembly (QIR) produced by the compiler. It is clear that \compilername{} is
highly effective for realistic benchmarks at inlining Qwerty programs into a
single QIR function suitable for the QIR Base Profile, but also that 
\compilername{} preserves soundness by still emitting callables when necessary (e.g.,
when inlining is disabled).

\begin{table}
\caption{QIR callable intrinsics generated by different compiler configurations.}\label{tab:callables}
\begin{tabular}{r|l|l|l|l|l|l}
\toprule
 & \multicolumn{2}{l|}{\textbf{Q\#}} & \multicolumn{2}{l|}{\textbf{\compilername{} (No Opt)}} & \multicolumn{2}{l}{\textbf{\compilername{} (Opt)}} \\
 & create & inv. & create & inv. & create & inv. \\ \midrule
B--V & 5 & 8 & 3 & 3 & \textbf{0} & \textbf{0} \\
D--J & 4 & 4 & 2 & 2 & \textbf{0} & \textbf{0} \\
Grover & 6 & 4 & 5 & 16 & \textbf{0} & \textbf{0} \\
Period & 12 & 16 & 4 & 4 & \textbf{0} & \textbf{0} \\
Simon & 4 & 4 & 7 & 7 & \textbf{0} & \textbf{0} \\
\bottomrule
\end{tabular}
\end{table}

\subsection{Comparison with Gate-Oriented Languages}\label{sec:eval-gates}
In this section, we consider the overhead of writing common quantum programs in
Qwerty and compiling with \compilername{} versus writing in traditional circuit-oriented languages and compiling with their respective compilers. Our methodology follows: (1) generate quantum assembly~\cite{cross_open_2017,cross_openqasm_2022} from all
five benchmarks in all four languages for different oracle input sizes; (2) optimize the resulting code with the Qiskit transpiler set to
\texttt{-O3}~\cite{Qiskit}; and (3) feed the resulting optimized assembly into
Azure Quantum Resource Estimator~\cite{van_dam_using_2023}, which estimates
\textbf{physical qubit count} and \textbf{runtime} for the circuit on fault-tolerant hardware.
Both metrics are important because functionally equivalent circuit designs often face a tradeoff of qubit count versus runtime~\cite{schmitt_tweedledum_2022,fowler_surface_2012}; increasing qubit count may make algorithms out of reach of hardware, yet a longer runtime may carry greater monetary cost and gives more opportunity for error to accumulate.
We use the default estimation parameters, which model a $[\![338,1,13]\!]$
surface code with a \SI{5.2}{\us} cycle time~\cite{van_dam_using_2023}. The Modern Q\# QDK compiler was used to compile Q\#~\cite{modern_qdk}.

Results are shown in Figs.~\ref{fig:eval-runtime} and \ref{fig:eval-qubits}. (Deutsch--Jozsa results, not shown, were virtually identical to Bernstein--Vazirani due to their similar circuit structure.)
The circuits generated by \compilername{} consistently keep pace with
circuit-orient\-ed languages despite the Qwerty programs containing no
hand-programmed gates. The Tweedledum-powered circuit synthesis in \compilername{}
(Section~\ref{sec:classical-synth}) consistently matches or beats the
performance of the oracle synthesis in the Quipper compiler, likely due to
Quipper's willingness to use ancilla qubits for XOR operations, which is
intentionally avoided by the synthesis algorithm implemented in
Tweedledum~\cite{schmitt_compilation_2021}.
The Q\# compiler and \compilername{} outperform other compilers significantly for Grover's
(Figs.~\ref{fig:eval-runtime:grover} and \ref{fig:eval-qubits:grover}) due to using Selinger's decomposition of multi-controlled gates (see Section~\ref{sec:quantum-opt})~\cite{selinger_quantum_2013}.
Note that small variations in performance, such as Quipper's slight deviations from others in period finding (Figs.~\ref{fig:eval-runtime:period} and \ref{fig:eval-qubits:period}) are caused by slightly differently circuit structure being marginally more or less favorable to the Qiskit transpiler --- in this case, this difference is Quipper using renaming-based swaps for IQFT rather than SWAP gates.

These results show that even with Qwerty's higher level of
abstraction, \compilername{} enables Qwerty programs to perform 
similarly to handwritten quantum circuits.

\section{Related Work}\label{sec:relwork}
\noindent\textbf{Quantum MLIR dialects:}
To our knowledge,
operations in the Qwerty MLIR dialect (Section~\ref{sec:qwerty-dialect})
can better signify programmer intent and program structure than most existing
quantum MLIR dialects; for example, the op to prepare a primitive basis state
(\lstinline[style=ir]!qbprep!) is distinct from the basis translation op used for state evolution (\lstinline[style=ir]!qbtrans!). Dialects
such as QIRO or QSSA would likely represent both cases with
low-level gate ops instead~\cite{ittah_qiro_2022,peduri_qssa_2022}.
Similarly, both CUDA-Q's \texttt{Quake}
dialect~\cite{The_CUDA-Q_development_team_CUDA-Q} and IBM's \texttt{QCS} and
\texttt{QUIR} consist of gates as the fundamental quantum operation on
qubits~\cite{healy_design_2024}.
In comparing levels of abstraction between the Qwerty MLIR dialect and prior quantum
dialects, Xanadu's \texttt{quantum} dialect in their Catalyst compiler is an
exception, however. While the dialect mostly contains lower level gate
operations, it does contain some higher abstraction operations, such as an
operation dedicated to computing a Hamiltonian observation given an array of
observables as input~\cite{Ittah_Catalyst_a_Python_2024}. This is a higher
level of abstraction than Qwerty reasons about, as it is specific to an
application domain (quantum chemistry).

\noindent\textbf{Quantum compilers:}
Prior quantum compilers have implemented similar features to our work, such as
auto-generation of function specializations (Section~\ref{sec:func-spec-gen})
and synthesis of circuits from a classical description
(Section~\ref{sec:classical-synth}). Q\#, for example, allows writing
\texttt{Adjoint f} and \texttt{Controlled f}, for a function value \texttt{f},
so Q\# compilers can auto-synthesize the specializations that these expressions
request~\cite{svore_q_2018,qsharp_specializations}. However, Qwerty's
predication (\lstinline!b & f!) is more general than a control in
circuit-oriented programming.
Additionally,
the compilers for Scaffold, Quipper, and Q\# can all automatically synthesize
quantum circuits from classical circuits, which can then be used as oracles and
subroutines in a programmer's quantum
program~\cite{abhari_scaffold_2012,javadiabhari_scaffcc_2014,qir_oracle_generation_22,green_quipper_2013}.

\section{Conclusion}
Qwerty is a high-level quantum programming language designed around
bases and functions instead of gates. This new quantum programming paradigm introduces new challenges in
compilation, including synthesizing circuits and specializing adjoint or
predicated forms of functions.
This paper presented \textit{\compilername{}}, a compiler that answers
these challenges with a type system and a series of high and low-level optimization passes.
\compilername{} uses a new, custom dialect of MLIR, \textit{Qwerty IR,}
which is capable of being lowered to a form ready
for both simulation and execution on near-term hardware.  In a comparison with the output
of circuit-oriented compilers on hand-coded programs, \compilername{} has been shown to produce high-quality code across a set of well-known quantum algorithms.
Future upgrades to \compilername{} should guarantee optimizations that are necessary for hardware targets with limited classical capability, such as inlining, and implement language features that support more physics-focused applications, such as Hamiltonian simulation.

\begin{acks}
The authors thank the anonymous reviewers and Pulkit Gupta for their helpful feedback.
We acknowledge support for this work
from NSF planning grant \#2016666
and through the ORNL STAQCS project.
Support for this work also came from the U.S. Department of Energy, Office of
Science, Advanced Scientific Computing Research Accelerated Research in
Quantum Computing Program under field work proposal ERKJ332.
This research was supported in part through research infrastructure and
services provided by the Rogues Gallery
testbed~\cite{powell_wrangling_2019,young_experimental_2019} hosted by the
Center for Research into Novel Computing Hierarchies (CRNCH) at Georgia Tech.
The Rogues Gallery testbed is primarily supported by the
NSF under NSF Award Number \#2016701.
\end{acks}

\appendix
\section{Artifact Appendix}

\subsection{Abstract}
Our artifact contains quantum algorithm benchmarks (written in Q\#, Quipper,
Qiskit, and Qwerty), scripts for extracting resource estimates from these
benchmarks, and the \compilername{} compiler and runtime for Qwerty. Only a
Linux machine with Docker and sufficient memory (approximately 32 GB of RAM) is
needed to run the evaluation.

\subsection{Artifact check-list (meta-information)}

{\small
\begin{itemize}
  \item {\bf Algorithm: } Quantum circuit synthesis
  \item {\bf Program: } Custom benchmarks, included in artifact
  \item {\bf Compilation: } gcc 11.4.0 and LLVM 19.1.2, both included in Docker image
  \item {\bf Transformations: } Custom compiler passes, included in artifact
  \item {\bf Binary: } x86\_64 binaries for Linux (Ubuntu 22.04.5 LTS) packaged as Docker image
  \item {\bf Metrics: } (1) number of calls to QIR intrinsics; (2) estimated physical qubits for a surface code; and (3) estimated runtime on a surface code
  \item {\bf Output: } Table and graphs (Table~\ref{tab:callables} and Figs.~\ref{fig:eval-runtime}-\ref{fig:eval-qubits}, respectively)
  \item {\bf Experiments: } The artifact README describes how to run experiments
  \item {\bf How much disk space required (approximately)?: } Approximately \SI{30}{\giga\byte} (evaluation results are only \SI{50}{\mega\byte}, but both Docker images are \SI{14}{\giga\byte} combined)
  \item {\bf How much time is needed to prepare workflow (approximately)?: } Under 1 hour to download and load Docker images
  \item {\bf How much time is needed to complete experiments (approximately)?: } Under 4 hours
  \item {\bf Publicly available?: } Yes
  \item {\bf Code licenses (if publicly available)?: } MIT license
  \item {\bf Archived (provide DOI)?: } 10.5281/zenodo.14505385
\end{itemize}}

\subsection{Description}

\subsubsection{How delivered}
The artifact is available on Zenodo at the following link: \url{https://doi.org/10.5281/zenodo.14505385}

\subsubsection{Hardware dependencies} Sufficient memory, approximately \SI{32}{\giga\byte}
total. Approximately \SI{30}{\giga\byte} of disk space.

\subsubsection{Software dependencies} Docker on Linux.

\subsection{Installation}

Run the following two commands:

\begin{verbatim}
docker image load -i \
    qwerty-artifact-docker.tar.xz
docker image load -i \
    qwerty-artifact-quipper-docker.tar.xz
\end{verbatim}

\subsection{Experiment workflow}

The following command will compile Quipper programs to quantum assembly (this
is done separately because Quipper requires a specific system configuration and
thus its own Docker image):

\begin{verbatim}
mkdir data
docker run --rm -v $(pwd)/data/:/data/ \
    qwerty-artifact-quipper quipper-bench-qasm.sh
\end{verbatim}

\noindent
(The generated assembly can be found on the host at \\
\texttt{data/quipper/*.qasm}.)

\noindent
The following command will run the evaluation pipeline:

\begin{verbatim}
docker run --rm -v $(pwd)/data:/data/ \
    qwerty-artifact eval/run-eval.sh
\end{verbatim}

\noindent
Afterward, \texttt{data/summary/} on the host contains a summary of the results
of the evaluation.

\noindent
Detailed results are found in other subdirectories of \texttt{data/}: the
statistics for QIR callables for Q\# versus Qwerty
(Section~\ref{sec:eval-inline}) can be found in
\texttt{data/count-callables/}, and detailed resource estimation
(Section~\ref{sec:eval-gates}) results are in \\
\texttt{data/compare-circs/}.

\subsection{Evaluation and expected result}

Table~\ref{tab:callables} should match \texttt{data/summary/table.csv} exactly. \\
Fig.~\ref{fig:eval-runtime} should match \texttt{data/summary/time\_*.pdf}.
Fig.~\ref{fig:eval-qubits} should match \texttt{data/summary/physical\_*.pdf}.

\subsection{Notes}
The full \compilername{} source code is included in the artifact as
\texttt{qwerty-artifact-source.tar.xz}. (This code is also available at
\texttt{/qwerty/} in the \texttt{qwerty-artifact} Docker image.)
Inside the source tarball (or at \texttt{/qwerty/} in the Docker image),
\texttt{README.md} explains how to build the compiler from source. Other files
in \texttt{docs/} document the code structure, testing, debugging, evaluation,
and integration with other tools.

% Add "B" to the start of algorithm numbers
% https://tex.stackexchange.com/a/554566/76494
\setcounter{algocf}{0}
\makeatletter
\renewcommand{\thealgocf}{\@Alph\c@section\arabic{algocf}}
\renewcommand{\thefigure}{\@Alph\c@section\arabic{figure}}
\makeatother

\appendix
\setcounter{section}{1}

\section{Span Checking Details}\label{app:span-checking}

The efficient span equivalence checking algorithm described in
Section~\ref{sec:span-equiv-checking}\ofmain{} is written as pseudocode in
Algorithm~\ref{alg:span-check}.
The algorithm proves span
equivalence using case analysis; its correctness can be proved inductively
using a similar proof by cases. Algorithm~\ref{alg:span-factor} elaborates on
sub-cases of the factoring performed on line~\ref{alg:span-check:factor} of
Algorithm~\ref{alg:span-check}. Lines
\ref{alg:span-factor:caseonestart}-\ref{alg:span-factor:caseoneend} of
Algorithm~\ref{alg:span-factor} hold because of Lemmas~\ref{lem:span1} and
\ref{lem:span2}.
(Below, $\hilbertspace{}$ denotes the single-qubit Hilbert space
$\vspan(\ket{0},\ket{1})$.)

\begin{algo}
\SetKwData{succ}{success}\SetKwData{fail}{failure}
\SetKwData{lq}{ldeque}\SetKwData{rq}{rdeque}
\SetKwData{left}{$\ell$}\SetKwData{right}{$r$}
\SetKwData{big}{big}\SetKwData{small}{small}

\Input{A basis translation $\varbin$\texttt{ >{}> }$\varbout$}
\Output{Either success or failure}
\BlankLine
\lq $\gets$ elements of $\varbin$, each normalized \;
\rq $\gets$ elements of $\varbout$, each normalized \;
\While{\lq and \rq are not empty}{
    \left $\gets$ pop front of \lq \;
    \right $\gets$ pop front of \rq \;
    \eIf{$\mathrm{dim}$ \left $=$ $\mathrm{dim}$ \right}{
        \eIf{\left $=$ \right, or \left and \right both fully span}{
            \Continue \;
        }{ % else
            \Return{\fail}
        }
    }{ % else
        \big $\gets \max(\ell,r)$ \;
        \small $\gets \min(\ell,r)$ \;
        \eIf{factoring \small from \big fails (Algorithm.~\ref{alg:span-factor}) \nllabel{alg:span-check:factor}}{
            \Return{\fail}
        }{ % else
            \Continue \;
        }
    }
}
\If{either \lq or \rq is not empty}{
    \Return{\fail}\algocomment{Dimension mismatch}
}
\Return{\succ}
\caption{Span equivalence checking for a basis translation.}\label{alg:span-check}
\end{algo}

\begin{algo}
\SetKwData{bigq}{bigdeque}
\SetKwData{succ}{success}\SetKwData{fail}{failure}
\SetKwData{big}{big}\SetKwData{small}{small}

\Input{Two basis elements, \big and \small, and the deque for \big from Algorithm~\ref{alg:span-check} (\bigq)}
\Output{Either success or failure}
\BlankLine
\uIf{both \big and \small fully span \nllabel{alg:span-factor:caseonestart}}{
    $\delta \gets (\mathrm{dim}\: \big) - (\mathrm{dim}\: \small)$ \;
    $\varprim \gets$ primitive basis of \big \;
    Push $\varprim$\lstinline![!$\delta$\lstinline!]! to front of \bigq \;
    \Return{\succ} \nllabel{alg:span-factor:caseoneend}
} \uElseIf{\small fully spans and \big is a basis literal \nllabel{alg:span-factor:casetwostart}} {
    \If{If factoring succeeds (Algorithm~\ref{alg:full-span-factor}) \nllabel{alg:span-factor:full-span-factor}} {
        Push remainder to front of \bigq \;
        \Return{\succ}
    }
} \ElseIf{both \small and \big are basis literals} {
    \If{If factoring succeeds (Algorithm~\ref{alg:vec-span-factor})} {
        Push remainder to front of \bigq \;
        \Return{\succ}
    }
} \nllabel{alg:span-factor:casetwoend}
\Return{\fail} \algocomment{Fallthrough}
\caption{Factoring a basis element in span checking.}\label{alg:span-factor}
\end{algo}

\begin{algo}
\SetKwData{bl}{bl}\SetKwData{nvar}{n}\SetKwData{mvar}{m}
\SetKwData{fnd}{found}\SetKwData{ctr}{ctr}\SetKwData{prefixes}{prefixes}
\SetKwData{succ}{success}\SetKwData{fail}{failure}

\Input{A basis literal $\bl = \texttt{\{}\mathrm{bv}_1\texttt{,}\mathrm{bv}_2\texttt{,}\ldots\texttt{,}\mathrm{bv}_{\mvar}\texttt{\}}$ (with $\mathrm{bv}_i$ sorted lexicographically) and a positive integer \nvar such that $(\mathrm{dim}\: \bl) > \nvar$}
\Output{Either success (and a remainder) or failure}
\BlankLine
\If{\mvar is not divisible by $2^{\nvar}$ \nllabel{alg:full-span-factor:divcheck}}{
    \Return{\fail}
}
Count distinct \nvar-bit prefixes across all $\mathrm{bv}_i$ \;
\If{there are less than $2^{\nvar}$ distinct prefixes}{
    \Return{\fail}
}
Count occurrences of $(\mathrm{dim}\:\bl-\nvar)$-bit suffixes across all $\mathrm{bv}_i$ \;
\If{any suffix appears less than $2^{\nvar}$ times}{
    \Return{\fail}
}
\Return{\succ and distinct suffixes}\algocomment{Remainder}
\caption{Factoring \lstinline!std[!\textsf{n}\lstinline!]!, \lstinline!pm[!\textsf{n}\lstinline!]!, or \lstinline!ij[!\textsf{n}\lstinline!]! from a basis literal \textsf{bl} in span checking. Bit operations are on eigenbits.}\label{alg:full-span-factor}
\end{algo}

\begin{algo}
\SetKwData{bl}{bl}\SetKwData{nvar}{n}\SetKwData{mvar}{m}
\SetKwData{fnd}{found}\SetKwData{ctr}{ctr}\SetKwData{prefixes}{prefixes}
\SetKwData{succ}{success}\SetKwData{fail}{failure}

\Input{Basis literals $\bl = \texttt{\{}\mathrm{bv}_1\texttt{,}\mathrm{bv}_2\texttt{,}\ldots\texttt{,}\mathrm{bv}_{\mvar}\texttt{\}}$ and $\bl' = \texttt{\{}\mathrm{bv}_1'\texttt{,}\mathrm{bv}_2'\texttt{,}\ldots\texttt{,}\mathrm{bv}_{\mvar'}'\texttt{\}}$, both sorted lexicographically and with $(\mathrm{dim}\: \bl) > (\mathrm{dim}\: \bl')$}
\Output{Either success (and a remainder) or failure}
\BlankLine
\If{\bl and $\bl'$ have different primitive bases \nllabel{alg:vec-span-factor:paulicheck}}{
    \Return{\fail}
}
\If{\mvar is not divisible by $\mvar'$ \nllabel{alg:vec-span-factor:divcheck}}{
    \Return{\fail}
}
$\nvar \gets \mathrm{dim}\:\bl'$ \;
Count distinct \nvar-bit prefixes across all $\mathrm{bv}_i$ \;
\If{there are less than $\mvar'$ distinct prefixes, or any prefix is not equal to some $\mathrm{bv}_i'$ \nllabel{alg:vec-span-factor:prefixcheck}}{
    \Return{\fail}
}
Count occurrences of $(\mathrm{dim}\:\bl-\nvar)$-bit suffixes across all $\mathrm{bv}_i$ \;
\If{any suffix appears less than $\mvar'$ times}{
    \Return{\fail}
}
\Return{\succ and distinct suffixes}\algocomment{Remainder}
\caption{Factoring a basis literal \textsf{bl}$'$ from a basis literal \textsf{bl} in span checking. Bit operations are on eigenbits.}\label{alg:vec-span-factor}
\end{algo}

\begin{lemma}\label{lem:span1}
$\vspan($\lstinline!fourier[!$N$\lstinline!]!$) =
\vspan($\lstinline!fourier[!$N_1$\lstinline!]!$) \otimes$ \\
$\vspan($\lstinline!fourier[!$N_2$\lstinline!]!$)$
if $N = N_1 + N_2$ and $N_1,N_2 \in \mathbb{Z}^{+}$.
\begin{proof}
Follows from $\hilbertspace^{\otimes N_1+N_2} = \hilbertspace^{\otimes N_1}\otimes\hilbertspace^{\otimes N_2}$.
\end{proof}
\end{lemma}
\begin{lemma}\label{lem:span2}
$\vspan($\lstinline!fourier[!$N$\lstinline!]!$) =
 \vspan($\lstinline!std[!$N$\lstinline!]!$) ={}$ \\
$\vspan($\lstinline!pm[!$N$\lstinline!]!$) =
 \vspan($\lstinline!ij[!$N$\lstinline!]!$) =
 \vspan(\texttt{\{}\mathrm{bv}_1\texttt{,}\mathrm{bv}_2\texttt{,}\ldots\texttt{,}\mathrm{bv}_m\texttt{\}})$ if the basis literal is well-typed, fully spans, and has dimension $N \in \mathbb{Z}^{+}$.
\begin{proof}
All spans are $\hilbertspace^{\otimes N}$.
\end{proof}
\end{lemma}
The remaining cases, lines \ref{alg:span-factor:casetwostart}-\ref{alg:span-factor:casetwoend} of
Algorithm~\ref{alg:span-factor}, hold due to Lemma~\ref{lem:factor}:
\begin{lemma}\label{lem:factor}
There exist $bv_j'$ such that
\[
    \texttt{\{}\mathrm{bv}_1\texttt{,}\mathrm{bv}_2\texttt{,}\ldots\texttt{,}\mathrm{bv}_{m_1}\texttt{\}} = \texttt{\{}\left(\left(\mathrm{bv}_{(k)}\mathrm{bv}_j'\right)_{j=1}^{m_2}\right)_{k=0}^{2^n-1}\texttt{\}}
\]
if and only if
\begin{align*}
    &\vspan(\texttt{\{}\mathrm{bv}_1\texttt{,}\mathrm{bv}_2\texttt{,}\ldots\texttt{,}\mathrm{bv}_{m_1}\texttt{\}}) \\
    &=\hilbertspace^{\otimes n} \otimes \vspan(\texttt{\{}\mathrm{bv}_1'\texttt{,}\mathrm{bv}_2'\texttt{,}\ldots\texttt{,}\mathrm{bv}_{m_2}'\texttt{\}}).
\end{align*}
(The notation $(\cdot)_i$ denotes the existence of some ordering of a comma-separated list. $\mathrm{bv}_{(k)}$ represents a basis vector with the same primitive basis as all $\mathrm{bv}_i$ and eigenbits corresponding to the $n$-bit binary representation of $k$.)
\begin{proof} \ \\
$(\bm{\rightarrow})$:  Definition of the tensor product on vector spaces. \\
$(\bm{\leftarrow})$:  Similar.
\end{proof}
\end{lemma}
Algorithm~\ref{alg:full-span-factor}, used on line~\ref{alg:span-factor:full-span-factor} of Algorithm~\ref{alg:span-factor}, implements Lemma~\ref{lem:factor}. It uses the following corollary as a short-circuit optimization (line \ref{alg:full-span-factor:divcheck} of Algorithm~\ref{alg:full-span-factor}):
\begin{corollary}\label{cor:divcheck}
If $2^n$ does not divide $m_1$, then there are no $\mathrm{bv}_j'$ such that
\begin{align*}
    &\vspan(\texttt{\{}\mathrm{bv}_1\texttt{,}\mathrm{bv}_2\texttt{,}\ldots\texttt{,}\mathrm{bv}_{m_1}\texttt{\}}) \\
    &= \hilbertspace^{\otimes n} \otimes \vspan(\texttt{\{}\mathrm{bv}_1'\texttt{,}\mathrm{bv}_2'\texttt{,}\ldots\texttt{,}\mathrm{bv}_{m_2}'\texttt{\}}).
\end{align*}
\begin{proof}
\textit{(By contraposition)} Suppose there exist $\mathrm{bv}_j'$
such that
\begin{align*}
    &\vspan(\texttt{\{}\mathrm{bv}_1\texttt{,}\mathrm{bv}_2\texttt{,}\ldots\texttt{,}\mathrm{bv}_{m_1}\texttt{\}}) \\
    &= \hilbertspace^{\otimes n} \otimes \vspan(\texttt{\{}\mathrm{bv}_1'\texttt{,}\mathrm{bv}_2'\texttt{,}\ldots\texttt{,}\mathrm{bv}_{m_2}'\texttt{\}}).
\end{align*}
By Lemma~\ref{lem:factor},
\[
    \texttt{\{}\mathrm{bv}_1\texttt{,}\mathrm{bv}_2\texttt{,}\ldots\texttt{,}\mathrm{bv}_{m_1}\texttt{\}} = \texttt{\{}\left(\left(\mathrm{bv}_{(k)}\mathrm{bv}_j'\right)_{j=1}^{m_2}\right)_{k=0}^{2^n-1}\texttt{\}}.
\]
Clearly,
\[
    \left\vert\texttt{\{}\left(\left(\mathrm{bv}_{(k)}\mathrm{bv}_j'\right)_{j=1}^{m_2}\right)_{k=0}^{2^n-1}\texttt{\}}\right\vert = 2^n \times m_2.
\]
Thus, $2^n \times m_2 = m_1$, i.e., $m_1$ is a multiple of $2^n$.
\end{proof}
\end{corollary}
Algorithm~\ref{alg:vec-span-factor} appeals to a straightforward generalization
of Lemma~\ref{lem:factor} and Corollary~\ref{cor:divcheck}
(Algorithm~\ref{alg:full-span-factor}) to factor out an arbitrary basis literal
rather than only a fully-spanning basis. This requires some additional checks, namely lines~\ref{alg:vec-span-factor:paulicheck} and \ref{alg:vec-span-factor:prefixcheck}
of Algorithm~\ref{alg:vec-span-factor}.

\begin{lemma}\label{lem:span-check-complexity}
Algorithms~\ref{alg:full-span-factor} and \ref{alg:vec-span-factor} run in time $O(k\log k)$ if $k$ is the total number of AST nodes in the inputs.
\begin{proof}
In Algorithm~\ref{alg:full-span-factor}, \textsf{bl} can be sorted twice, first by prefixes and second by suffixes, to perform the counting needed.
Algorithm~\ref{alg:vec-span-factor} is similar: $\textsf{bl}'$ can be sorted as well and binary searched to detect unexpected prefixes.
\end{proof}
\end{lemma}

\begin{theorem}\label{thm:span-check-time}
Span equivalence (Algorithm~\ref{alg:span-check}) runs in time $O(k^2\log k)$ if $k$ is the total number of AST nodes in the basis translation.
\begin{proof}
In the worst case, each of the $O(k)$ iterations of Algorithm~\ref{alg:span-check} requires factoring, which is $O(k\log k)$ per Lemma \ref{lem:span-check-complexity}.
\end{proof}
\end{theorem}

\section{Inlining Conditionals in Qwerty IR}\label{app:inlining-cond}
For an example of the inlining technique described at the end of
Section~\ref{sec:inlining}\ofmain{}, consider lines~\lnteleportevilbegin{} and
\lnteleportevilend{} of Fig.~\ref{fig:teleport}. After lifting lambdas to
functions, the IR produced by line~\lnteleportevilbegin{} of
Fig.~\ref{fig:teleport} resembles the following:

\begin{lstlisting}[style=irs]
%23 = scf.if %22 {
  %28 = qwerty.func_const @lambda3[]
  scf.yield %28
} else {
  %28 = qwerty.func_const @lambda4[]
  scf.yield %28
}
%24 = qwerty.call_indirect %23(%10)
\end{lstlisting}

Clearly, this does not resemble the canonicalization pattern of a
\lstinline[style=ir]!call_indirect! with a \lstinline[style=ir]!func_const!
operand mentioned in Section~\ref{sec:inlining}\ofmain{} because there is an
\lstinline[style=ir]!scf.if! in the way. To address this, \compilername{}
includes a specialized canonicalization pattern that transforms this IR into
the following:

\begin{lstlisting}[style=irs]
%24 = scf.if %22 {
  %28 = qwerty.func_const @lambda3[]
  %29 = qwerty.call_indirect %28(%10)
  scf.yield %29
} else {
  %28 = qwerty.func_const @lambda4[]
  %29 = qwerty.call_indirect %28(%10)
  scf.yield %29
}
\end{lstlisting}

Subsequently, the desired pattern is recognized and the indirect calls shown
are converted to direct calls. Similar patterns are also implemented for the
case when the result of an \lstinline[style=ir]!scf.if! is passed to a
\lstinline[style=ir]!func_adj! or \lstinline[style=ir]!func_pred!. Note that a
similar transformation would be more difficult to implement in the presence of
unstructured control flow (e.g., in QIR).

\begin{figure}
\centering
\begin{lstlisting}[style=numxs,framexleftmargin=4mm,xleftmargin=8mm,linewidth=0.95\linewidth]
@qpu
def teleport(secret: qubit) -> qubit:
    alice, bob = 'p0' | '1' & std.flip

    m_pm, m_std = \
        secret + alice | '1' & std.flip \
                       | (pm + std).measure

    secret_teleported = \
        bob | (pm.flip if m_std else id) \
            | (std.flip if m_pm else id)

    return secret_teleported
\end{lstlisting}
\caption{Quantum teleportation in Qwerty\label{fig:teleport}. The syntax \lstinline!std.flip! is syntactic sugar for \lstinline!std >> \{'1','0'\}!. The semantics of \lstinline!pm.flip! is similar.}
\end{figure}

\section{Function Specialization Details}\label{app:specgen}
Section~\ref{sec:func-spec-gen}\ofmain{} describes how \compilername{} determines set of
necessary function specializations, and Algorithm~\ref{alg:specgen} shows this
process as pseudocode. Because cyclic call graphs are not possible to construct
in Qwerty, Algorithm~\ref{alg:specgen} will always terminate.

\begin{algo}
\Input{Qwerty IR after function specialization analysis}
\Output{Set of specializations $(\mathrm{funcName},\mathrm{isAdj},\mathrm{numCtrls})$ to generate}
\BlankLine
$V, E \gets \emptyset$ \;
\For{every function $f$}{
    $V \gets V \cup \{(\text{name of }f, \texttt{false}, 0)\}$ \;
}
\For{every \texttt{call} and \texttt{call\_indirect} op}{
    \For{every callee $v'$}{
        $V \gets V \cup \{v'\}$ \;
        $E \gets E \cup \{((\text{name of caller}, \texttt{false}, 0), v')\}$ \;
    }
}
\Repeat{$V$ and $E$ stop changing}{
    \For{every function $f$}{
        $v \gets (\text{name of }f, \texttt{false}, 0)$ \;
        \For{every edge $(v,v') \in E$}{
            $(\mathrm{funcName}',\mathrm{isAdj}',\mathrm{numCtrls}') \gets v'$ \;
            \For{every $u \in V$ with name of $f$}{
                $(\mathrm{funcName},\mathrm{isAdj},\mathrm{numCtrls}) \gets u$ \;
                $v'' \gets (\mathrm{funcName}',\; \mathrm{isAdj} \oplus \mathrm{isAdj}',$ \hspace{-1mm} \;
                $\qquad\quad\mathrm{numCtrls} + \mathrm{numCtrls}')$ \;
                $V \gets V \cup \{v''\}$ \;
                $E \gets E \cup \{(u, v'')\}$ \;
            }
        }
    }
}
Run DFS from entry point and mark nodes reached \;
Delete unmarked nodes from $V$ \;
\Return{$V$}
\caption{Determining function specializations needed to execute some Qwerty IR}\label{alg:specgen}
\end{algo}

\section{Standardization Details}\label{app:standardize}

\begin{figure}
\begin{adjustbox}{width=0.87\linewidth}\begin{minipage}{\linewidth}\centering\input{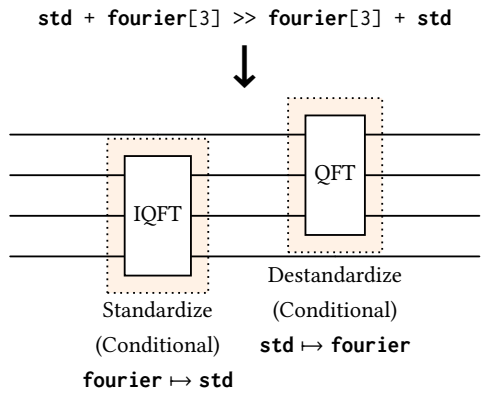}\end{minipage}\end{adjustbox}
\Description{A figure showing how a basis translation is compiled with QFT}
\caption{Compiling a basis translation with inseparable (de)standardizations}\label{fig:synth-qft}
\end{figure}

\begin{algo}
\SetKwData{lstd}{lstd}\SetKwData{rstd}{rstd}
\SetKwData{lq}{ldeque}\SetKwData{rq}{rdeque}
\SetKwData{left}{$\ell$}\SetKwData{right}{$r$}
\SetKwData{big}{big}\SetKwData{small}{small}
\SetKwData{bigq}{bigdeque}
\SetKwData{bigstd}{bigstd}\SetKwData{smallstd}{smallstd}
\SetKwData{kind}{kind}

\Input{A well-typed basis translation $\varbin$\texttt{ >{}> }$\varbout$}
\Output{A list of standardizations and a list of destandardizations, where each element is a tuple $(\mathrm{primBasis},\mathrm{dim},\mathrm{isConditional})$}
\BlankLine
\lstd, \rstd $\gets$ $[]$ \;
\lq $\gets$ elements of $\varbin$ \;
\rq $\gets$ elements of $\varbout$ \;
\While{\lq and \rq are not empty}{
    \left $\gets$ pop front of \lq \;
    \right $\gets$ pop front of \rq \;
    \eIf{neither \left nor \right are padding and $\vprim \left = \vprim \right$}{
        \kind $\gets$ Unconditional \;
    }{ % else
        \kind $\gets$ Conditional \;
    }
    \eIf{$\vdim$ \left $=$ $\vdim$ \right}{
        \If{\left is not padding}{
            append $(\vprim \left, \vdim \left, \kind)$ to \lstd
        }
        \If{\right is not padding}{
            append $(\vprim \right, \vdim \right, \kind)$ to \rstd
        }
    }{ % else
        \big $\gets \max(\ell,r)$ \;
        \small $\gets \min(\ell,r)$ \;
        $\delta \gets (\vdim \big) - (\vdim \small)$ \;

        \eIf{$\vprim \big$ is separable and not padding}{
            \If{\small is not padding}{
                append $(\vprim \small,\vdim \small,\kind)$ to \smallstd \;
            }
            append $(\vprim \big, \vdim \small, \kind)$ to \bigstd \;
            push $(\vprim \big)[\delta]$ to front of \bigq \;
        }{ % else
            \If{\small is not padding}{
                append $(\vprim \small, \vdim \small,$ $\text{Conditional})$ to \smallstd \;
            }
            \If{\big is not padding}{
                append $(\vprim \big, \vdim \big,$ $\text{Conditional})$ to \bigstd \;
            }
            push padding$[\delta]$ to front of \bigq \;
        }
    }
}
\Return{\lstd and \rstd}

\caption{Determining standardizations for a basis translation}\label{alg:standardize}
\end{algo}

Section~\ref{sec:lower-btrans}\ofmain{} gives some basic insight into the process of determining standardizations, but in realistic Qwerty compilation,
the presence of \textit{inseparable} primitive bases (a basis that cannot be
written as a tensor product of smaller bases) is a serious complication.
For example, even though both three-qubit Fourier bases
overlap in the basis translation
\lstinline!std + fourier[3] >> fourier[3] + std!,
there are no unconditional (de)stand\-ard\-izations. Instead, the compiler must
synthesize a circuit (Fig.~\ref{fig:synth-qft}) invokes the inverse quantum Fourier
transform (IQFT) on the rightmost three qubits and then the quantum Fourier
transform (QFT) on the leftmost three qubits. It would be invalid to try to run
IQFT only on the rightmost qubit and then QFT only on the leftmost because the
Fourier basis is inseparable.

Algorithm~\ref{alg:standardize} handles the situation of standardizing in the
presence of inseparable bases while also maintaining the invariant that the
head of both deques start at the same qubit by delicately inserting padding basis
elements when inseparable basis elements are encountered.

\begin{algo}
\SetKwData{succ}{success}\SetKwData{fail}{failure}
\SetKwData{lq}{ldeque}\SetKwData{rq}{rdeque}
\SetKwData{leftl}{$\ell$}\SetKwData{rightr}{$r$}
\SetKwData{big}{big}\SetKwData{small}{small}
\SetKwData{bigq}{bigdeque}
\SetKwData{factor}{factor}

\Input{A well-typed basis translation $\varbin$\texttt{ >{}> }$\varbout$}
\Output{An aligned basis translation $\varbin'$\texttt{ >{}> }$\varbout'$}
\BlankLine
$\varbin',\:\varbout' \gets$ empty bases \;
\lq $\gets$ elements of $\varbin$, each standardized \nllabel{alg:align:qinit1} \;
\rq $\gets$ elements of $\varbout$, each standardized \nllabel{alg:align:qinit2} \;
\While{\lq and \rq are not empty}{
    \leftl $\gets$ pop front of \lq \;
    \rightr $\gets$ pop front of \rq \;
    \eIf{$\vdim\leftl = \vdim\rightr$}{
        \uIf{$\leftl ={}$\lstinline!std[N]! and $\rightr \ne{}$\lstinline!std[N]!}{
            $\leftl \gets {}$ \lstinline!std[N]! as a basis literal
        } \ElseIf{$\leftl \ne{}$\lstinline!std[N]! and $\rightr ={}$\lstinline!std[N]!}{
            $\rightr \gets {}$ \lstinline!std[N]! as a basis literal
        }
        $\varbin' \gets \varbin' \toktensor \leftl$ \;
        $\varbout' \gets \varbout' \toktensor \rightr$ \;
    }{ % else
        \big $\gets \max(\ell,r)$ \;
        \small $\gets \min(\ell,r)$ \;
        \uIf{$\big ={}$ \lstinline!std[N]! \nllabel{alg:align:factor1}}{
            $\factor \gets{}$\lstinline!std[!$\vdim \small$\lstinline!]! \;
            \If{$\small \ne{}$\lstinline!std[N]!}{
                $\factor \gets {}$ $\factor$ as a basis literal \;
            }
            $\varbbig' \gets \varbbig' \toktensor \factor$ \;
            $\varbsmall' \gets \varbsmall' \toktensor \small$ \;
            $\delta \gets (\vdim \big) - (\vdim \small)$ \;
            Push \lstinline!std[!$\delta$\lstinline!]! to front of \bigq \;
        } \uElseIf{a $(\vdim \small)$-dimensional basis literal $\small'$ can be factored from \big \nllabel{alg:align:factor2}}{
            \If{$\small ={}$\lstinline!std[N]!}{
                $\small \gets {}$ $\small$ as a basis literal \;
            }
            $\varbbig' \gets \varbbig' \toktensor \small'$ \;
            $\varbsmall' \gets \varbsmall' \toktensor \small$ \;
            Push remainder to front of \bigq \;
        } \Else {
            Merge until $\big'$ and $\small'$ are basis elements of equal dimension \nllabel{alg:align:merge} \;
            $\varbbig' \gets \varbbig' \toktensor \big'$ \;
            $\varbsmall' \gets \varbsmall' \toktensor \small'$ \;
        }
    }
}
\Return{$\varbin'$\texttt{ >{}> }$\varbout'$}
\caption{Aligning basis translations}\label{alg:align}
\end{algo}

\section{Basis Alignment Details}\label{app:align}
The permutation step of basis translation lowering as described in
Section~\ref{sec:lower-btrans}\ofmain{} can encounter some ambiguities.
For example, consider the following basis translation:
\begin{center}
\lstinline!{'1'} + std >> {'11','10'}!
\end{center}
The compiler must decide whether to interpret this as
\begin{center}
\lstinline!{'1'} + {'0','1'} >> {'1'} + {'1','0'}!
\end{center}
or
\begin{center}
\lstinline!{'10','11'} >> {'11','10'}!.
\end{center}
For the former, \compilername{} would synthesize a permutation that swaps
$\ket{0}$ and $\ket{1}$ and then add a control afterward to honor the
\lstinline!{'1'}! predicate. The latter would instead synthesize a permutation that
swaps $\ket{10}$ and $\ket{11}$.

\textit{Alignment} addresses this ambiguity by producing a functionally
equivalent basis translation $\varbin'$\texttt{ >{}> }$\varbout'$ such that two
conditions are met for each elementwise pair of basis elements $\varbein$ and
$\varbeout$: first, $(\vdim \varbein) = (\vdim \varbeout)$; and second,
$\varbein$ is a basis literal if and only if $\varbeout$ is a basis literal.
Both new basis translations in the previous paragraph are aligned forms of the
first. However, factoring (the first aligned form above) is preferred over
merging (the second aligned form) because a factored form is more structured;
for instance, it can reduce the size of the permutation to synthesize.
Factoring may not always be possible, however. For example, imagine the basis
translation
\begin{center}
\lstinline!{'0','1'} + {'0','1'} >> {'00','10','01','11'}!.
\end{center}
The right-hand side cannot be written as the tensor product of two basis
literals, so factoring is not possible. In such cases, the compiler resorts to
merging (e.g., merging
\discretionary{\lstinline!{'0','1'} +!}{\lstinline!{'0','1'}!}%
              {\lstinline!{'0','1'} + {'0','1'}!}
on the left-hand side of the previous example to form
\discretionary{\lstinline!{'00','01','10',!}{\lstinline!'11'}!}{\lstinline!{'00','01','10','11'}!}).

Algorithm~\ref{alg:align} shows the procedure for aligning a basis translation. It attempts to factor if
possible (lines~\ref{alg:align:factor1} and \ref{alg:align:factor2}), falling
back to merging if necessary (line~\ref{alg:align:merge}).
\textit{Standardizing} a basis element, mentioned on on lines
\ref{alg:align:qinit1} and \ref{alg:align:qinit2}, means changing its primitive
basis to \lstinline!std! and removing all vector phases.

\section{Modifications Needed in QIR}\label{app:qir-changes}
When compiling to QIR, different function specializations are lowered to different LLVM
functions. In accordance with the QIR callables API~\cite{qir_callables}, a
struct is generated for captures, and a table in static memory holds pointers to
the different specializations of a given Qwerty function. The only modification to QIR
needed by Qwerty is to modify the
\discretionary{\texttt{\_\_quantum\_\_rt\_\_callable}}{\texttt{\_invoke}}
              {\texttt{\_\_quantum\_\_rt\_\_callable\_invoke}}
intrinsic not to mangle arguments passed
to the callable. In Q\#, if \lstinline!f! has type
$\mathrm{Qubit}[] \rightarrow \mathrm{Unit}$, then \lstinline!Controlled f! has
type $(\mathrm{Qubit}[],${\allowbreak}$\mathrm{Qubit}[]) \rightarrow \mathrm{Unit}$, whereas
in Qwerty, if \lstinline!f! has type $\mathrm{Qubit}[N]${\allowbreak}$\rightarrow
\mathrm{Qubit}[N]$, then \lstinline!b & f! has type $\mathrm{Qubit}[M+N]
\rightarrow \mathrm{Qubit}[${\allowbreak}$M+N]$ if $\vdim \texttt{b} = M$~\cite{adams_qwerty_2024}. Thus, multiple uses
of \:\lstinline!&!\: for the same function in Qwerty only increase the size of the
qubit register argument, whereas in Q\#, multiple uses of \texttt{Controlled}
add more qubit arrays to the input tuple. In this case, QIR is hard-coded for
Q\# and attempts to collapse many qubit array arguments into one qubit array of
controls, which is incorrect for QIR generated by \compilername{}.

\bibliographystyle{ACM-Reference-Format}
\bibliography{qwerty-compiler}

\end{document}
\endinput